\documentclass[english,10pt, aps, prd, a4paper, preprintnumbers,floatfix, nofootinbib, showpacs, superscriptaddress, notitlepage]{revtex4-1}
\usepackage{graphicx}
\usepackage{rotate}
\usepackage{amsmath}
\usepackage{amssymb}
\usepackage{float}
\usepackage{comment}
\usepackage[normalem]{ulem}
\usepackage[usenames,dvipsnames]{color}
\usepackage[colorinlistoftodos]{todonotes}
\usepackage{array}
\usepackage{amsmath}
\usepackage{amssymb}
\usepackage[colorlinks=true,citecolor=darkred,urlcolor=darkred, pdfborder={0 0 0}]{hyperref}
\usepackage{cancel}

\pdfoutput=1
\newcommand {\ignore}[1]{}
\definecolor{darkred}{rgb}{0.6,0,0}

\def\SM{$\mathrm{SU(3)_c \otimes SU(2)_L \otimes U(1)_Y}$}
\def\tt1{$\mathrm{SU(3) \otimes SU(3)_L \otimes U(1)}$}
\def\0331{$\mathrm{SU(3)_{c} \otimes SU(3)_L \otimes U(1) }$}
\def\3311{$\mathrm{SU(3)_{c} \otimes SU(3)_L \otimes U(1)_X \otimes U(1)_{N}}$}
\def\gsim{\raise0.3ex\hbox{$\;>$\kern-0.75em\raise-1.1ex\hbox{$\sim\;$}}}
\def\lsim{\raise0.3ex\hbox{$\;<$\kern-0.75em\raise-1.1ex\hbox{$\sim\;$}}}

\definecolor{mightnightblue}{RGB}{25,25,112}
\definecolor{brown}{rgb}{0.59, 0.29, 0.0}
\newcommand{\dgreen}{\color[rgb]{0,0.5,0}}

\def\vev#1{\left\langle #1\right\rangle}

\def\21{$\mathrm{SU(2)_L \otimes U(1)_Y}$}

\newcommand{\AddrAHEP}{  AHEP Group, Institut de F\'{i}sica Corpuscular --
  CSIC/Universitat de Val\`{e}ncia, Parc Cient\'ific de Paterna.\\
 C/ Catedr\'atico Jos\'e Beltr\'an, 2 E-46980 Paterna (Valencia) - SPAIN}
\newcommand{\AddrUSM}{Universidad T\'{e}cnica Federico Santa Mar\'{\i}a, Casilla 110-V, Valpara\'{\i}so, Chile}
\newcommand{\CCTVAL}{Centro Cient\'{\i}fico-Tecnol\'ogico de Valpara\'{\i}so, Casilla 110-V, Valpara\'{\i}so, Chile}
\newcommand{\SAPHIR}{Millennium Institute for Subatomic Physics at High-Energy Frontier (SAPHIR), Fern\'andez Concha 700, Santiago, Chile}
\newcommand{\AddrUNAB}{Departamento de Ciencias F\'isicas, Universidad Andres Bello, \\
   Sazi\'e 2212, Piso 7, Santiago, Chile}
\newcommand{\AddrUM}{Physik Department T70, Technische Universit\"at M\"unchen,\\
James-Franck-Stra{\ss}e 1, D-85748 Garching, Germany}
\input{tcilatex}
\begin{document}

\title{\color{BrickRed} Scotogenic neutrino masses with gauged matter parity and gauge coupling unification}
\author{A. E. C\'{a}rcamo Hern\'{a}ndez}
\email{antonio.carcamo@usm.cl}
\affiliation{\AddrUSM}
\affiliation{\CCTVAL}
\affiliation{\SAPHIR}
\author{Chandan Hati}
\email{c.hati@tum.de }
\affiliation{\AddrUM}
\author{Sergey Kovalenko}
\email{sergey.kovalenko@unab.cl}
\affiliation{\AddrUNAB}
\affiliation{\CCTVAL}
\affiliation{\SAPHIR}
\author{Jos\'{e} W. F. Valle}
\email{valle@ific.uv.es}
\affiliation{\AddrAHEP}
\author{Carlos A. Vaquera-Araujo}
\email{vaquera@fisica.ugto.mx}
\affiliation{Departamento de F\'isica, DCI, Campus Le\'on, Universidad de Guanajuato, Loma del Bosque 103, Lomas del Campestre C.P. 37150, Le\'on, Guanajuato, M\'exico}
\affiliation{Consejo Nacional de Ciencia y Tecnolog\'ia, Avenida Insurgentes Sur 1582. Colonia Cr\'edito Constructor, Alcald\'ia Benito Ju\'arez, C.P. 03940, Ciudad de M\'exico, M\'exico}
\affiliation{Dual CP Institute of High Energy Physics, C.P. 28045, Colima, M\'exico}
\date{\today }

\begin{abstract}
  \vspace{1cm}

  Building up on previous work we propose a Dark Matter (DM) model with gauged matter parity and dynamical gauge coupling unification,
  driven by the same physics responsible for scotogenic neutrino mass generation.
  Our construction is based on the extended gauge group \3311, whose spontaneous breaking leaves a residual conserved matter parity, $M_{P}$, stabilizing the DM particle candidates of the model.    
  A key role is played by the Majorana ${\rm SU(3)_{L}}$-octet leptons, in allowing successful gauge coupling unification and one-loop scotogenic neutrino mass generation.
  Theoretical consistency allows for a \emph{plethora} of new particles at the $\lsim \mathcal{O}$(10) TeV scale, hence accessible to future collider and low-energy experiments.
  
\end{abstract}

\maketitle

\affiliation{\AddrUSM}

\affiliation{\AddrUM}

\affiliation{\AddrUNAB}

\affiliation{\AddrAHEP}

\affiliation{Departamento de F\'isica, DCI, Campus Le\'on, Universidad de Guanajuato, Loma del Bosque 103, Lomas del Campestre C.P. 37150, Le\'on, Guanajuato, M\'exico}

\affiliation{Consejo Nacional de Ciencia y Tecnolog\'ia, Avenida Insurgentes Sur 1582. Colonia Cr\'edito Constructor, Alcald\'ia Benito Ju\'arez, C.P. 03940, Ciudad de M\'exico, M\'exico}





\noindent


\section{Introduction}
\label{sec:introduction} 

The supersymmetric approach to gauge coupling unification has so far not been vindicated experimentally, 
neither at colliders, nor through the observation of proton decay~\cite{ParticleDataGroup:2020ssz}.
However, the historical discovery of neutrino oscillations and the growing evidence for a weakly interacting massive particle (WIMP) as, perhaps, the most viable candidate for cosmological Dark Matter 
motivate us to seek new roads to unification. We start from the phenomenologically safe foundations provided by the Standard Model (SM) with the gauge group \SM~extended to \3311.
The use of $\mathrm{SU(3)_L}$ as an extended electroweak symmetry has been advocated by its ability to explain the observed number of families through
the anomaly cancellation requirement~\cite{Singer:1980sw,Frampton:1992wt}. 
Moreover, a suggestion was made within the original \0331 framework proposed in~\cite{Singer:1980sw} of how neutrino masses and gauge coupling unification could emerge together~\cite{Boucenna:2014dia},
so that that the physics responsible for small neutrino masses could also drive the unification of the gauge couplings.
This model provides a radiative seesaw mechanism for calculable neutrino masses, arising from quantum corrections mediated by new \0331 gauge bosons.
Apart from its somewhat \emph{ad hoc} nature, the model lacked an explanation for cosmological dark matter, another major drawback of particle physics. 
The key property of any dark matter candidate is its stability on cosmological time scales, suggesting the existence of a (nearly) preserved stabilizing symmetry.
Recently it has been proposed that the latter could be a discrete residual matter parity symmetry, surviving the spontaneous breaking of the extended gauge symmetry~\cite{Dong:2014wsa,Dong:2015yra,Alves:2016fqe}.
This happens within a simple U(1)-extension of the gauge symmetry, allowing the implementation of a conserved matter parity 
\begin{equation}
  \label{eq:mp}
M_{P}=(-1)^{3(B-L)+2s}.
\end{equation}
analogous to the R-parity in supersymmetric theories. 

In this paper we put together all these attractive features, proposing a theory of calculable scotogenic Majorana neutrino masses~\cite{Ma:2006km} in the \3311 framework, or 3-3-1-1, for short.
Dark matter is a weakly interacting massive particle (WIMP) that mediates neutrino mass generation.
Gauge couplings unify for a 3-3-1 scale just above the TeV range, making the model directly testable at the LHC.
For that we introduce an $\mathrm{SU(3)_L}$ Majorana octet, responsible for both neutrino mass generation and gauge coupling unification.\\[-.4cm]

The paper is organized as follows.
In Sec.~\ref{sec:model-setup} we present the the model setup, while in Sec.~\ref{sec:scalar-sector} we discuss the scalar sector and symmetry breaking.
In Sec.~\ref{sec:yukawa sector} we analyse the neutrino mass matrix and the scotogenic mechanism.
Dark matter and gauge coupling unification are discussed in Secs.~\ref{sec:wimp-dark-matter} and ~\ref{sec:gauge-coupl-unif}, respectively.
A brief outlook is given in Sec.~\ref{sec:Conclusions}.

\section{The model setup}
\label{sec:model-setup}

\subsection{Field content}

We study a 3-3-1-1 model based on the \3311 gauge symmetry introduced in \cite{Alves:2016fqe} and the implementation of a conserved matter parity according to Eq.~(\ref{eq:mp}).
This is very much analogous to the R-parity symmetry imposed on the supersymmetric theories.
Except that the $M_{P}$ discrete symmetry is not imposed \emph{ad hoc} as a global symmetry, but rather arises as a remnant of the spontaneously broken $B-L$ gauge group \cite{Dong:2014wsa,Dong:2015yra,Alves:2016fqe}.
It leads to the stability of the lightest $M_{P}$-odd particle, and hence to a potentially viable WIMP dark matter candidate.
In our model, electric charge $Q$ and the $B-L$ generators are embedded into the gauge symmetry as 
\begin{align}
Q & = T_3-\frac{T_8}{\sqrt{3}}+X,\label{Q} \\
B-L & = -\frac{2}{\sqrt{3}}T_8+N\label{BminusL},
\end{align}
with $T_i$ $(i = 1,2,3,...,8)$, $X$ and $N$ being the respective generators of $\mathrm{SU(3)_L}$, $\mathrm{U(1)_X}$ and $\mathrm{U(1)_N}$.
The lepton sector of the model includes three leptonic triplets\footnote{Notice that in the original 3-3-1 formulation  \cite{Singer:1980sw}, left-handed leptons transform as anti-triplets of $\mathrm{SU(3)_L}$. This configuration can be recovered by exchanging all triplets and anti-triplets in the present work.}
\begin{equation}
l_{aL}=%
\begin{pmatrix}
\nu _{a} \\ 
e_{a} \\ 
N_{a}%
\end{pmatrix}%
_{L},
\end{equation}
each with its new neutral lepton $N_{aL}$, $a=1,2,3$.
The quark fields are arranged as 
\begin{equation}
q_{iL}=%
\begin{pmatrix}
d_{i} \\ 
-u_{i} \\ 
D_{i}%
\end{pmatrix}%
_{L}\qquad q_{3L}=%
\begin{pmatrix}
u_{3} \\ 
d_{3} \\ 
U_{3}%
\end{pmatrix}%
_{L}.
\end{equation}%
Anomaly cancellation requires that two families of quarks $q_{iL}$; $i=1,2$ transform as anti-triplets and one $q_{3L}$ as a triplet, as in the original Singer-Valle-Schechter (SVS) scheme~\cite{Singer:1980sw}.
This way, the number of the fermion families is equal to the number of colors, thus allowing a natural explanation of the number of generations of  the SM fermions.
The full field content is shown in Table \ref{Table:Field-Content}. 
  \begin{table}[th]
\centering\label{tab:tab3311} 
\begin{tabular}{|c|c|c|c|c|c|c|}
\hline
\hspace{0.2cm} Field \hspace{0.2cm} & \hspace{0.2cm}$\mathrm{SU(3)_{c}}$ \hspace{0.2cm%
} & \hspace{0.2cm} $\mathrm{SU(3)_{L}}$ \hspace{0.2cm} & \hspace{0.2cm}$\mathrm{U(1)_{X}}$ 
\hspace{0.2cm} & \hspace{0.2cm}$\mathrm{U(1)_{N}}$ \hspace{0.2cm} & \hspace{0.2cm} $Q$%
\hspace{0.2cm} & \hspace{0.2cm} $M_{P}=(-1)^{3(B-L)+2s}$ \hspace{0.2cm} \\ 
\hline\hline
$q_{iL}$ & \textbf{3} & $\overline{{\mathbf{3}}}$ & 0 & 0 & $(-\frac{1}{3},%
\frac{2}{3},-\frac{1}{3})^{T}$ & $(++-)^{T}$ \\ 
$q_{3L}$ & \textbf{3} & \textbf{3} & $\frac{1}{3}$ & $\frac{2}{3}$ & $(\frac{%
2}{3},-\frac{1}{3},\frac{2}{3})^{T}$ & $(++-)^{T}$ \\ 
$u_{aR}$ & \textbf{3} & \textbf{1} & $\frac{2}{3}$ & $\frac{1}{3}$ & $\frac{2%
}{3}$ & $+$ \\ 
$d_{aR}$ & \textbf{3} & \textbf{1} & $-\frac{1}{3}$ & $\frac{1}{3}$ & $-%
\frac{1}{3}$ & $+$ \\ 
$U_{3R}$ & \textbf{3} & \textbf{1} & $\frac{2}{3}$ & $\frac{4}{3}$ & $\frac{2%
}{3}$ & $-$ \\ 
$D_{iR}$ & \textbf{3} & \textbf{1} & $-\frac{1}{3}$ & $-\frac{2}{3}$ & $-%
\frac{1}{3}$ & $-$ \\ 
$l_{aL}$ & \textbf{1} & \textbf{3} & $-\frac{1}{3}$ & $-\frac{2}{3}$ & $%
(0,-1,0)^{T}$ & $(++-)^{T}$ \\ 
$e_{aR}$ & \textbf{1} & \textbf{1} & $-1$ & $-1$ & $-1$ & $+$ \\ \hline
$\nu _{iR}$ & \textbf{1} & \textbf{1} & $0$ & $-4$ & $0$ & $-$ \\ 
$\nu _{3R}$ & \textbf{1} & \textbf{1} & $0$ & $5$ & $0$ & $+$ \\ 
$\Omega _{aL}$ & \textbf{1} & \textbf{8} & $0$ & $0$ & $\left( 
\begin{matrix}
0 & 1 & 0 \\ 
-1 & 0 & -1 \\ 
0 & 1 & 0%
\end{matrix}%
\right) $ & $\left( 
\begin{matrix}
- & - & + \\ 
- & - & + \\ 
+ & + & -%
\end{matrix}%
\right) $ \\ \hline\hline
$\eta $ & \textbf{1} & \textbf{3} & $-\frac{1}{3}$ & $\frac{1}{3}$ & $%
(0,-1,0)^{T}$ & $(++-)^{T}$ \\ 
$\rho $ & \textbf{1} & \textbf{3} & $\frac{2}{3}$ & $\frac{1}{3}$ & $%
(1,0,1)^{T}$ & $(++-)^{T}$ \\ 
$\chi $ & \textbf{1} & \textbf{3} & $-\frac{1}{3}$ & $-\frac{2}{3}$ & $%
(0,-1,0)^{T}$ & $(--+)^{T}$ \\ 
$\phi $ & \textbf{1} & \textbf{1} & $0$ & $2$ & $0$ & $+$ \\ \hline
$\sigma $ & \textbf{1} & \textbf{1} & $0$ & $1$ & $0$ & $-$ \\ \hline
\end{tabular}%
\caption{3311 model field content ($a=1,2,3$ and $i=1,2$ are family indices). Note the non-standard charges of the $\protect\nu _{R}$.}
\label{Table:Field-Content}
\end{table}

Notice that, in addition to the field content of the model of Ref.~\cite{Alves:2016fqe}, our present setup includes three Majorana octets $\Omega_{aL}$ ($a=1,2,3$).
They are crucial for achieving a successful gauge coupling unification scenario, as well as for providing the tiny masses of the light active neutrinos via a one-loop scotogenic mechanism.
The latter is possible thanks to the inclusion of a scalar singlet $\sigma$. 
This electrically neutral field has a nontrivial charge under the preserved remnant matter-parity symmetry, providing a viable scalar WIMP dark matter candidate. 
Our model also includes three right-handed neutrinos with non-standard $\mathrm{U(1)_N}$ charges, which have been introduced in \cite{CarcamoHernandez:2020ehn} to ensure an anomaly free gauge symmetry. 
These fields do not take part in the neutrino mass generation mechanism, and they do not mix with the other neutral fermions of the model.
The two $\nu _{iR}$ fermion fields can acquire a Majorana mass after spontaneous symmetry breaking, by the inclusion of a scalar field transforming as $(\mathbf{1},\mathbf{1},0,8)$,
 while a mass term for $\nu _{3R}$ requires a scalar with quantum numbers $(\mathbf{1},\mathbf{1},0,-10)$. 
 In order to keep the analysis of the scalar sector as simple as possible here we do not include those extra scalar fields. \\[-.4cm]

  The gauged $B-L$ symmetry is spontaneously broken by two units as the singlet scalar $\phi$ develops a vacuum expectation value (VEV).
  As seen from the assignments in Table~\ref{Table:Field-Content} this leaves a discrete remnant symmetry $M_{P}$ specified in Eq.(\ref{eq:mp}). 
%
  The most general VEV alignment for the scalars consistent with a preserved $M_P$ symmetry is~\cite{VanDong:2018yae}
\begin{equation}
\langle\eta\rangle=\frac{1}{\sqrt{2}}(v_1,0,0)^T,\quad \langle\rho\rangle=%
\frac{1}{\sqrt{2}}(0,v_2,0)^T, \quad\langle\chi\rangle=\frac{1}{\sqrt{2}}(0,0,w)^T, \quad
\langle\phi\rangle=\frac{1}{\sqrt{2}}\Lambda,\quad \langle\sigma\rangle=0.
\end{equation}
Here we will assume the hierarchy $w, \Lambda,\gg v_1, v_2 $, leading to the following spontaneous symmetry breaking (SSB) pattern 
\begin{align}
\mathrm{SU(3)_{C}}\times &\mathrm{SU(3)_{L}}\times \mathrm{U(1)_{X}}\times \mathrm{U(1)_{N}}  \notag
\\
&\downarrow w, \Lambda  \notag \\
\mathrm{SU(3)_C} &\times \mathrm{SU(2)_{L}}\times \mathrm{U(1)_{Y}}\times M_P^{}  \notag \\
\label{eq:SSB-chain}
&\downarrow v_1,v_2  \notag \\
\mathrm{SU(3)_C} & \times \mathrm{U(1)_{Q}}\times M_P^{}\,.
\end{align}

\subsection{Majorana octet lepton fields}

The presence of the $\mathrm{SU(3)_L}$ octet fermions, $\Omega$, is a key ingredient of the model in order to ensure gauge coupling unification and to generate radiative masses for the active neutrinos. 
In this subsection we will show the explicit derivation of the electric charge and matter parity assignments of the components of the Majorana leptonic octets.
Our starting point is the $\mathrm{SU(3)_L}$ algebra, which is described by  
\begin{eqnarray}  \label{eq:alg-1}
[T_{a},T_{b}]&=& i f_{abc}T_{c},\ \ \ \ T_{a}=\frac{\lambda_{a}}{2},
\end{eqnarray}
where $\lambda_{a}$, $a=1,\dots,8$ are the Gell-Mann matrices. 
The corresponding Cartan subalgebra is spanned by $H_{1}= T_{3}$ and $H_{2}=T_{8}$. The $\mathrm{SU(3)_L}$ algebra in the Cartan basis reads 
\begin{eqnarray}  \label{eq:alg-cart-1}
[H_{i}, E_{\alpha_a}] &=& (\alpha_a)_{i} E_{\alpha_a}, \ \ \ E^{\dagger}_{\alpha_a} =
E_{-\alpha_a},\qquad i=1,2,\,\,a=1,2,3,
\end{eqnarray}
where the ladder operators $E_{\alpha_a}$ are given by
\begin{eqnarray}  \label{eq:ladder-2}
&&E_{\alpha_{1}} = \frac{1}{\sqrt{2}} \left(T_{4} + i T_{5}\right) = \frac{1}{%
\sqrt{2}} 
\begin{pmatrix}
0 & 0 & 1 \\ 
0 & 0 & 0 \\ 
0 & 0 & 0%
\end{pmatrix}%
, \\
&&E_{\alpha_{2}}  = \frac{1}{\sqrt{2}} \left(T_{6} - i T_{7}\right)= \frac{1}{%
\sqrt{2}} 
\begin{pmatrix}
0 & 0 & 0 \\ 
0 & 0 & 0 \\ 
0 & 1 & 0%
\end{pmatrix}%
, \\
&&E_{\alpha_{3}}  = \frac{1}{\sqrt{2}} \left(T_{1} + i T_{2}\right) = \frac{1}{%
\sqrt{2}} 
\begin{pmatrix}
0 & 1 & 0 \\ 
0 & 0 & 0 \\ 
0 & 0 & 0%
\end{pmatrix}%
,
\end{eqnarray}
and the coefficients $(\alpha_a)_{i}$ are the components of the roots 
\begin{eqnarray}  \label{eq:roots-1}
\vec{\alpha}_{1}&=&\left(\frac{1}{2}, \frac{\sqrt{3}}{2}\right),\hspace{8mm} 
\vec{\alpha}_{2}=\left(\frac{1}{2}, -\frac{\sqrt{3}}{2}\right), \hspace{8mm} 
\vec{\alpha}_{3}=(1,0).
\end{eqnarray}
The $\mathrm{SU(3)_L}$ leptonic octet, $\Omega$, can be decomposed in the Cartan basis of the $\mathrm{SU(3)_L}$ generators as follows  
\begin{eqnarray}  \label{eq:octet-decomp-1}
\Omega&=& \Omega^{(3)} T_{3}+\Omega^{(8)} T_{8} + \Omega^{(1)}_{\pm} E_{\pm
\alpha_{1}}+\Omega^{(2)}_{\pm} E_{\pm
\alpha_{2}}+ \Omega^{(3)}_{\pm} E_{\pm
\alpha_{3}}.
\end{eqnarray}
Using Eqs. (\ref{eq:mp}), (\ref{Q}), (\ref{BminusL}) and the charge assignments $Q_{X}(\Omega) = Q_{N}(\Omega) =0$ given in Table~\ref{Table:Field-Content},
we find the $Q$ and $B-L$ charge assignments as well as matter parity $M_{P}$ of the component fields in (\ref{eq:octet-decomp-1}), namely 
\begin{eqnarray}  \label{eq:Omega-assignments-2}
&&\Omega(T_{3}, T_{8}): \hspace{14mm} \Omega^{(3)}(0, 0), \ \
\Omega^{(8)}(0,0), \ \ \Omega^{(1)}_{\pm}\left(\pm \frac{1}{2}, \pm\frac{%
\sqrt{3}}{2}\right),\ \ \Omega^{(2)}_{\pm}\left(\pm \frac{1}{2}, \mp\frac{%
\sqrt{3}}{2}\right),\ \ \Omega^{(3)}_{\pm}\left(\pm 1, 0\right) \\
\label{eq:Omega-assignments-2-1}
&&\Omega(Q, B-L; M_{P}): \ \ \Omega^{(3)}(0, 0;-1), \ \
\Omega^{(8)}(0,0;-1), \ \ \Omega^{(1)}_{\pm}\left(0,\mp 1; +1\right),\ \
\Omega^{(2)}_{\pm}\left(\pm 1,\pm 1; +1\right),\ \
\Omega^{(3)}_{\pm}\left(\pm 1,0;-1\right).
\end{eqnarray}
Then, the leptonic Octet can be expressed in matrix form as 
\begin{eqnarray}  \label{eq:Octet-matrix-1}
&&\Omega_{L} = \frac{1}{\sqrt{2}} 
\begin{pmatrix}
\frac{1}{\sqrt{2}}\Omega^{(3)} +\frac{1}{\sqrt{6}}\ \Omega^{(8)} & 
\Omega^{(3)}_{+} & \Omega^{(1)}_{+} \\ 
&  &  \\ 
\Omega^{(3)}_{-} & -\frac{1}{\sqrt{2}}\Omega^{(3)} +\frac{1}{\sqrt{6}}\
\Omega^{(8)} & \Omega^{(2)}_{-} \\ 
&  &  \\ 
\Omega^{(1)}_{-} & \Omega^{(2)}_{+} & -\frac{2}{\sqrt{6}}\ \Omega^{(8)}%
\end{pmatrix}%
_{L},
\end{eqnarray}
Note that $(\Omega^{c})_{ij}= (\Omega_{ji})^{c}$. Therefore, $Q(\Omega_{ij})=Q(\Omega ^{c}_{\,ij})$ and $M_{P}(\Omega_{ij})=M_{P}(\Omega ^{c}_{\, ij})$,
as should be for a Majorana field $\Omega_{M} = \Omega_{L} \oplus (\Omega_{L})^{c}$.
From Eqs.~(\ref{eq:Omega-assignments-2-1}) and (\ref{eq:Octet-matrix-1}) we find the charge $Q$ and $M_{P}$ assignments of the leptonic octet, $\Omega$, shown in Table~\ref{Table:Field-Content}.   
%

In the following discussion, we will denote the components of $\Omega $ in canonical normalization as $\Omega ^{(3)}=\sqrt{2}\Psi ^{0}$, $%
\Omega ^{(8)}=\sqrt{2}\tilde{N}$, $\Omega _{\pm }^{(3)}=\sqrt{2}E^{\pm }$, $\Omega _{\pm }^{(2)}=\sqrt{2}\tilde{E}^{\pm }$, $\Omega _{+}^{(1)}=\sqrt{2}{\Delta }$, $\Omega _{-}^{(1)}=%
\sqrt{2}\tilde{\Delta}$. In this notation, the octet takes the form 
\begin{eqnarray}
&&\Omega _{L}=%
\begin{pmatrix}
\frac{1}{\sqrt{2}}\Psi +\frac{1}{\sqrt{6}}\ \tilde{N} & E^{+} & \Delta \\ 
&  &  \\ 
E^{-} & -\frac{1}{\sqrt{2}}\Psi +\frac{1}{\sqrt{6}}\ \tilde{N} & \tilde{E}%
^{-} \\ 
&  &  \\ 
\tilde{\Delta} & \tilde{E}^{+} & -\frac{2}{\sqrt{6}}\ \tilde{N}%
\end{pmatrix}%
_{L},  \label{eq:Octet-matrix-21} \\[5mm]
&&\left( \Omega _{L}\right) ^{c}=%
\begin{pmatrix}
\frac{1}{\sqrt{2}}(\Psi _{L})^{c}+\frac{1}{\sqrt{6}}\ (\tilde{N}_{L})^{c} & 
(E_{L}^{-})^{c} & (\tilde{\Delta}_{L})^{c} \\ 
&  &  \\ 
(E_{L}^{+})^{c} & -\frac{1}{\sqrt{2}}(\Psi _{L})^{c}+\frac{1}{\sqrt{6}}\ (%
\tilde{N}_{L})^{c} & (\tilde{E}_{L}^{+})^{c} \\ 
&  &  \\ 
({\Delta }_{L})^{c} & (\tilde{E}_{L}^{-})^{c} & -\frac{2}{\sqrt{6}}\ (\tilde{%
N}_{L})^{c}%
\end{pmatrix},\\[5mm]
&&\overline{\Omega }_{L}=\begin{pmatrix}
\frac{1}{\sqrt{2}}\overline{\Psi _{L}}+\frac{1}{\sqrt{6}}\ \overline{\tilde{N%
}_{L}} & \overline{E_{L}^{-}} & \overline{\tilde{\Delta}_{L}} \\ 
&  &  \\ 
\overline{E_{L}^{+}} & -\frac{1}{\sqrt{2}}\overline{\Psi _{L}}+\frac{1}{%
\sqrt{6}}\ \overline{\tilde{N}_{L}} & \overline{\tilde{E}_{L}^{+}} \\ 
&  &  \\ 
\overline{\Delta_{L}} & \overline{\tilde{E}_{L}^{-}} & -\frac{2}{%
\sqrt{6}}\ \overline{\tilde{N}_{L}}
\end{pmatrix}.
\end{eqnarray}

\section{Symmetry breaking}
\label{sec:scalar-sector}

The most general 3-3-1-1 gauge-invariant scalar potential of the model is given by
\begin{align}
\begin{split}
\label{eq:scalar-pot}
V=& \mu _{1}^{2}\rho ^{\dagger }\rho +\mu _{2}^{2}\chi ^{\dagger }\chi +\mu
_{3}^{2}\eta ^{\dagger }\eta +\mu _{4}^{2}\phi ^{\dagger }\phi +\mu
_{5}^{2}\sigma ^{\dagger }\sigma \\
& +\lambda _{1}(\rho ^{\dagger }\rho )^{2}+\lambda _{2}(\chi ^{\dagger }\chi
)^{2}+\lambda _{3}(\eta ^{\dagger }\eta )^{2} \\
& +\lambda _{4}(\rho ^{\dagger }\rho )(\chi ^{\dagger }\chi )+\lambda
_{5}(\rho ^{\dagger }\rho )(\eta ^{\dagger }\eta )+\lambda _{6}(\chi
^{\dagger }\chi )(\eta ^{\dagger }\eta ) \\
& +\lambda _{7}(\rho ^{\dagger }\chi )(\chi ^{\dagger }\rho )+\lambda
_{8}(\rho ^{\dagger }\eta )(\eta ^{\dagger }\rho )+\lambda _{9}(\chi
^{\dagger }\eta )(\eta ^{\dagger }\chi ) \\
& +\lambda _{10}(\phi ^{\dagger }\phi )(\rho ^{\dagger }\rho )+\lambda
_{11}(\phi ^{\dagger }\phi )(\chi ^{\dagger }\chi )+\lambda _{12}(\phi
^{\dagger }\phi )(\eta ^{\dagger }\eta ) \\
& +\lambda _{13}(\sigma ^{\dagger }\sigma )(\rho ^{\dagger }\rho )+\lambda
_{14}(\sigma ^{\dagger }\sigma )(\chi ^{\dagger }\chi )+\lambda _{15}(\sigma
^{\dagger }\sigma )(\eta ^{\dagger }\eta ) \\
& +\lambda _{16}(\phi ^{\dagger }\phi )^{2}+\lambda _{17}(\sigma ^{\dagger
}\sigma )^{2}+\lambda _{18}(\phi ^{\dagger }\phi )(\sigma ^{\dagger }\sigma
)+\lambda _{19}\left[ (\sigma ^{\dagger }\phi )(\eta ^{\dagger }\chi )+\text{%
h.c.}\right] \\
& +\left[ \frac{\mu _{t}}{2}\rho \eta \chi +\frac{\mu _{s}}{2}\phi ^{\dagger
}\sigma \sigma +\frac{\mu _{u}}{2}(\eta ^{\dagger }\chi )\sigma +\text{h.c.}%
\right], \ 
\end{split}%
\end{align}
where the $\lambda _{k}$ ($k=1,2,\cdots ,19$) are dimensionless parameters, whereas the $\mu _{r}$ ($r=1,2,\cdots ,5$), $\mu _{t}$, $\mu _{s}$, $\mu _{u} $ have dimension of mass.
We emphasize that in our model we have no need of imposing any global symmetries, all the ingredients for the scotogenic neutrino mass generation are already contained in the gauge symmetry group. 
To ensure that the $M_{P}$ symmetry remains conserved we require $\mu _{5}^{2}>0$, implying that the $M_{P}$-odd scalar $\sigma$ does not develop a nonzero VEV. 

The minimization conditions of the scalar potential yield the following relations:
\begin{eqnarray}
\mu _{1}^{2} &=&\frac{v_{1}w\mu _{t}-v_{2}\left( \lambda _{10}\Lambda
^{2}+2\lambda _{1}v_{2}^{2}+\lambda _{5}v_{1}^{2}+\lambda _{4}w^{2}\right) }{%
2v_{2}},  \notag \\
\mu _{2}^{2} &=&\frac{v_{1}v_{2}\mu _{t}-w\left( \lambda _{11}\Lambda
^{2}+\lambda _{4}v_{2}^{2}+\lambda _{6}v_{1}^{2}+2\lambda _{2}w^{2}\right) }{%
2w},  \notag \\
\mu _{3}^{2} &=&\frac{v_{2}w\mu _{t}-v_{1}\left( \lambda _{12}\Lambda
^{2}+2\lambda _{3}v_{1}^{2}+\lambda _{5}v_{2}^{2}+\lambda _{6}w^{2}\right) }{%
2v_{1}},  \notag \\
\mu _{4}^{2} &=&-\frac{1}{2}\left( 2\lambda _{16}\Lambda ^{2}+\lambda
_{10}v_{2}^{2}+\lambda _{12}v_{1}^{2}+\lambda _{11}w^{2}\right) .
\end{eqnarray}%

The scalar potential of the model has been previoulsy analyzed in Ref.~\cite{CarcamoHernandez:2020ehn}.
Here we just quote the results relevant for the modified neutrino mass generation mechanism of the present model.
Decomposing the scalar multiplets in components as
\begin{equation}
\label{scalar3plets3}
\eta=\left(
\begin{array}{c}
\frac{v_1+s_1+i a_1}{\sqrt{2}}\\
\eta_2^{-}\\
\frac{s'_3+i a'_3}{\sqrt{2}}
\end{array}
\right),\quad
\rho=\left(
\begin{array}{c}
\rho_1^{+}\\
\frac{v_2+s_2+i a_2}{\sqrt{2}}\\
\rho_3^{+}
\end{array}
\right),\quad
\chi=\left(
\begin{array}{c}
\frac{s'_1+i a'_1}{\sqrt{2}} \\
 \chi_2^{-} \\
 \frac{w+s_3+i a_3}{\sqrt{2}} \\
\end{array}
\right),\quad
\phi=\frac{\Lambda+s_{\phi}+i a_{\phi}}{\sqrt{2}},\quad
\sigma=\frac{s_{\sigma}+i a_{\sigma}}{\sqrt{2}}.
\end{equation}
The CP-even neutral scalar sector of the model was studied in \cite{CarcamoHernandez:2020ehn}. Besides the $125$ GeV SM-like Higgs identified with
\begin{equation}
h\approx\frac{v_1s_1+v_2s_2}{\sqrt{v_1^2+v_2^2}},
\end{equation}
and having a mass
\begin{equation}
m^2_h \approx \frac{\Lambda ^2 v_1 v_2 w \mu _t
\left(\left(\left(\lambda _4 \lambda _{11}-2 \lambda _2 \lambda _{10}\right) \lambda _{12}-\lambda _5 \left(\lambda _{11}^2-4 \lambda _2\lambda _{16}\right)+\lambda _6 \left(\lambda _{10} \lambda _{11}-2 \lambda _4 \lambda _{16}\right)\right) w^2-\lambda _{16} \mu _t^2\right)}{m^2_{H_1}m^2_{H_2}m^2_{H_3}}.
\end{equation}
There are three additional heavy Higgs bosons,  given by the approximate expressions
\begin{equation}
\begin{split}
H_1&\approx\frac{v_2s_1-v_1s_2}{\sqrt{v_1^2+v_2^2}},\qquad m^2_{H_1}\approx \frac{\left(v_1^2+v_2^2\right) w \mu _t}{2 v_1 v_2},\\
H_2&\approx\cos\xi s_3-\sin\xi s_4,\qquad m^2_{H_2}\approx \lambda _{16} \Lambda ^2+\lambda _2 w^2-\sqrt{\lambda _{16}^2 \Lambda ^4+\lambda _2^2 w^4+\lambda _{11}^2 \Lambda ^2 w^2-2 \lambda _2 \lambda _{16} \Lambda ^2 w^2},\\
H_3&\approx\sin\xi s_3+\cos\xi s_4,\qquad m^2_{H_3}\approx \lambda _{16} \Lambda ^2+\lambda _2 w^2+\sqrt{\lambda _{16}^2 \Lambda ^4+\lambda _2^2 w^4+\lambda _{11}^2 \Lambda ^2 w^2-2 \lambda _2 \lambda _{16} \Lambda ^2 w^2},
\end{split}
\end{equation}
valid under the assumption $\Lambda,w,\mu_t\gg v_1, v_2$. 
There are also two physical real scalars $\varphi _{1}$, $\varphi _{2}$ and one Nambu-Goldstone boson $G_1$, defined as
\begin{equation}\label{G1}
\left( 
\begin{array}{c}
\varphi_1 \\
\varphi_2 \\
G_1
\end{array}%
\right)=U^s \left( 
\begin{array}{c}
s'_1 \\
s'_3 \\
s_\sigma
\end{array}%
\right)=\left(
\begin{array}{ccc}
 \frac{v_1\cos\theta _s }{\sqrt{w^2+v_1^2}} & \frac{w \cos\theta _s}{\sqrt{w^2+v_1^2}} & \sin \theta _s\\
 -\frac{v_1\sin\theta _s }{\sqrt{w^2+v_1^2}} & -\frac{w \sin \theta _s}{\sqrt{w^2+v_1^2}} & \cos \theta _s \\
 \frac{w}{\sqrt{w^2+v_1^2}} & -\frac{v_1}{\sqrt{w^2+v_1^2}} & 0 \\
\end{array}
\right) \left( 
\begin{array}{c}
s'_1 \\
s'_3 \\
s_\sigma
\end{array}%
\right),
\end{equation}
where the mixing angle $\theta_s$ satisfies the relation
\begin{equation}\label{theta_s}
\tan2\theta_s=\frac{2 v_1 w \sqrt{v_1^2+w^2} \left(\lambda _{19} \Lambda +\mu _u\right)}{v_1 w \left(-2 \mu _5^2-\Lambda  \left(\lambda _{18} \Lambda +2 \mu _s\right)-\lambda
   _{13} v_2^2-\lambda _{15} v_1^2+\lambda _9 \left(v_1^2+w^2\right)-\lambda _{14} w^2\right)+v_2 \mu _t \left(v_1^2+w^2\right)}.
\end{equation}
The emergence of a Nambu-Goldstone boson in the CP-even scalar sector follows from of the existence of a non-hermitian gauge boson $X^0$,
  whose real part must absorb $G_1$ after the SSB, so as to acquire a consistent mass, while its imaginary part absorbs an analogous CP-odd Goldstone boson, as we discuss below.

The CP-odd neutral sector consists of four Nambu-Goldstone bosons, $G_{2,3,4,5}$, and three massive states, denoted as $A_1$, $\widetilde{\varphi}_1$ and $\widetilde{\varphi}_2$.
Three of these four Nambu-Goldstone bosons are given by  
\begin{equation}
G_2 =\frac{v_1a_1-v_2a_2}{\sqrt{v_1^2+v_2^2}},\qquad
G_{3}=\frac{v_1a_1-w a_2}{\sqrt{v_1^2+w^2}},\qquad
G_{4}=a_\phi,
\\
\end{equation}
and correspond to the longitudinal components of the physical  gauge bosons, $Z$, $Z^{\prime }$, $Z^{\prime \prime }$, respectively.
On the other hand, the massive state $A_1$ is 
\begin{equation}
A_1=\frac{v_2 w a_1+v_1 wa_2+v_1 v_2a_3}{\sqrt{(v_2 w)^2+(v_1 w)^2+(v_1 v_2)^2}},\qquad m^2_{A_1}= \frac{\mu _t \left(v_1^2 w^2+v_2^2 w^2+v_2^2 v_1^2\right)}{2 v_1 v_2 w},
\end{equation}
whereas the two physical states $\widetilde{\varphi}_1$, $\widetilde{\varphi}_2$ and the fifth Goldstone $G_5$ are defined as
\begin{equation}
\left( 
\begin{array}{c}
\widetilde{\varphi}_1 \\
\widetilde{\varphi}_2 \\
G_5
\end{array}%
\right)=U^a \left( 
\begin{array}{c}
a'_1 \\
a'_3 \\
a_\sigma
\end{array}%
\right)=\left(
\begin{array}{ccc}
 -\frac{v_1\cos \theta _a }{\sqrt{w^2+v_1^2}} & \frac{w \cos \theta _a}{\sqrt{w^2+v_1^2}} & \sin \theta _a \\
 \frac{v_1\sin \theta _a }{\sqrt{w^2+v_1^2}} & -\frac{w \sin \theta _a}{\sqrt{w^2+v_1^2}} & \cos \theta _a \\
 \frac{w}{\sqrt{w^2+v_1^2}} & \frac{v_1}{\sqrt{w^2+v_1^2}} & 0 \\
\end{array}
\right) \left( 
\begin{array}{c}
a'_1 \\
a'_3 \\
a_\sigma
\end{array}%
\right),
\end{equation}
with mixing angle
\begin{equation}
\tan2\theta_a=\frac{2 v_1 w \sqrt{v_1^2+w^2} \left(\mu _u-\lambda _{19} \Lambda \right)}{v_1 w \left(-\lambda _{18} \Lambda ^2-2 \mu _5^2+2 \Lambda  \mu _s-\lambda _{13}
   v_2^2-\lambda _{15} v_1^2+\lambda _9 \left(v_1^2+w^2\right)-\lambda _{14} w^2\right)+v_2 \mu _t \left(v_1^2+w^2\right)}.
\end{equation}

Notice that the Goldstone bosons $G_5$ and $G_{1}$ combine into a single complex neutral would-be Goldstone,  absorbed by the longitudinal component of the non-Hermitian neutral gauge boson $X^0$.

The real scalars $\varphi_1$, $\varphi_2$, $\widetilde{\varphi}_1$ and $\widetilde{\varphi}_2$ acquire squared masses given by
\begin{equation}
\begin{split}
&m_{\varphi_{1,2}}^2=\frac{1}{4 v_1 w}\Bigg\{v_1 w \left(\lambda _{18} \Lambda ^2+2 \mu _5^2+2 \Lambda  \mu _s+\lambda _{13} v_2^2+\lambda _{15} v_1^2+\lambda _9 \left(v_1^2+w^2\right)+\lambda _{14}
   w^2\right)+v_2 \mu _t \left(v_1^2+w^2\right)\\&\mp\mathcal{F}_s\Big\{\left(v_1 w \left(\lambda _{18} \Lambda ^2+2 \mu _5^2+2 \Lambda  \mu _s+\lambda _{13} v_2^2+\lambda _{15} v_1^2+\lambda _9
   \left(v_1^2+w^2\right)+\lambda _{14} w^2\right)+v_2 \mu _t \left(v_1^2+w^2\right)\right){}^2\\&-4 v_1 w \left(v_1^2+w^2\right) \big(v_2 \mu _t \left(\lambda _{18}
   \Lambda ^2+2 \mu _5^2+2 \Lambda  \mu _s+\lambda _{13} v_2^2+\lambda _{14} w^2\right)+v_1 w \big(\lambda _9 \left(\lambda _{18} \Lambda ^2+2 \mu _5^2+2 \Lambda 
   \mu _s+\lambda _{14} w^2\right)\\&-\left(\lambda _{19} \Lambda +\mu _u\right){}^2+\lambda _9 \lambda _{13} v_2^2\big)+\lambda _{15} v_2 v_1^2 \mu _t+\lambda _9
   \lambda _{15} v_1^3 w\big)\Big\}^{1/2}\Bigg\},\\
& \mathcal{F}_s=\mathrm{sgn}\left\{v_1 w \left(-2 \mu _5^2-\Lambda  \left(\lambda _{18} \Lambda +2 \mu _s\right)-\lambda
   _{13} v_2^2-\lambda _{15} v_1^2+\lambda _9 \left(v_1^2+w^2\right)-\lambda _{14} w^2\right)+v_2 \mu _t \left(v_1^2+w^2\right)\right\} ,\\
&m_{\widetilde{\varphi}_{1,2}}^2=\frac{1}{4 v_1 w}\Bigg\{v_1 w \left(\lambda _{18} \Lambda ^2+2 \mu _5^2-2 \Lambda  \mu _s+\lambda _{13} v_2^2+\lambda _{15} v_1^2+\lambda _9 \left(v_1^2+w^2\right)+\lambda _{14}
   w^2\right)+v_2 \mu _t \left(v_1^2+w^2\right)\\&\mp\mathcal{F}_a\Big\{\left(v_1 w \left(\lambda _{18} \Lambda ^2+2 \mu _5^2-2 \Lambda  \mu _s+\lambda _{13} v_2^2+\lambda _{15} v_1^2+\lambda _9 \left(v_1^2+w^2\right)+\lambda _{14}
   w^2\right)+v_2 \mu _t \left(v_1^2+w^2\right)\right){}^2\\&-4 v_1 w \left(v_1^2+w^2\right) \big(v_2 \mu _t \left(\lambda _{18} \Lambda ^2+2 \mu _5^2-2 \Lambda  \mu _s+\lambda _{13} v_2^2+\lambda _{14} w^2\right)+v_1 w \big(\lambda _9 \left(\lambda _{18} \Lambda ^2+2 \mu _5^2-2 \Lambda  \mu _s+\lambda _{14} w^2\right)\\&-\left(\lambda _{19} \Lambda-\mu _u \right){}^2+\lambda _9 \lambda _{13} v_2^2\big)+\lambda _{15} v_2 v_1^2 \mu _t+\lambda _9
   \lambda _{15} v_1^3 w\big)\Big\}^{1/2}\Bigg\},\\
& \mathcal{F}_a=\mathrm{sgn}\left\{v_1 w \left(-\lambda _{18} \Lambda ^2-2 \mu _5^2+2 \Lambda  \mu _s-\lambda _{13}
   v_2^2-\lambda _{15} v_1^2+\lambda _9 \left(v_1^2+w^2\right)-\lambda _{14} w^2\right)+v_2 \mu _t \left(v_1^2+w^2\right)\right\} , 
   \end{split}
 \end{equation}
 
 \vskip .2cm 
As shown in detail in Ref.~\cite{CarcamoHernandez:2020ehn}, in the limit $\mu _{s},~\mu_{u} \to 0$, one obtains a degenerate physical scalar spectrum $m_{\varphi _{1,2}}^{2}=m_{\widetilde{\varphi }_{1,2}}^{2}$.
 This degeneracy is lifted in our present model by the inclusion of the scalar singlet $\sigma $, a feature which is crucial in order to implement our scotogenic neutrino mass generation mechanism,
 as described in the next section.
\section{Yukawa sector}
\label{sec:yukawa sector}

The 3-3-1-1 gauge-invariant Lagrangian of the model responsible for the fermion mass terms is given by 
\begin{eqnarray}
  -\mathcal{L}_{\text{Yukawa}}&=&y_{3a}^{u}\overline{q}_{3L}\eta u_{aR}+y_{ia}^{u}\overline{q}_{iL}\rho
^{\ast }u_{aR}+y^{U}\overline{q}_{3L}\chi U_{3R}\notag \\
&&+y_{3a}^{d}\overline{q}_{3L}\rho d_{aR}+y_{ia}^{d}\overline{q}_{iL}\eta
^{\ast }d_{aR}+y_{ij}^{D}\overline{q}_{iL}\chi ^{\ast }D_{jR}\notag \\
&&+y_{ab}^{e}\overline{l}_{aL}\rho e_{bR} +\frac{M_{8\,ab}}{2}\mathrm{Tr}[\overline{\Omega }_{aL}(\Omega _{bL})^{c}]+y_{ab}^{\Omega }%
\overline{l}_{aL}(\Omega _{bL})^{c}\chi +\mathrm{h.c.}
\label{Ly}
\end{eqnarray}
It includes Yukawa interaction terms as well as the bare octet mass term. 
First note that the quark spectrum of our model coincides with the one presented in \cite{CarcamoHernandez:2020ehn}. 

We now focus on the tree-level masses of the neutral fermions. 
After spontaneous symmetry breaking, the last term in Eq. (\ref{Ly}) becomes 
\begin{equation}
\begin{split}
y_{ab}^{\Omega }%
\overline{l}_{aL}(\Omega _{bL})^{c}\chi=y_{ab}^{\Omega }\begin{pmatrix}
\overline{\nu}&\overline{e}&\overline{N}
\end{pmatrix}_{aL}
\begin{pmatrix}
\frac{1}{\sqrt{2}}(\Psi _{L})^{c}+\frac{1}{\sqrt{6}}\ (\tilde{N}_{L})^{c} & 
(E_{L}^{-})^{c} & (\tilde{\Delta}_{L})^{c} \\ 
&  &  \\ 
(E_{L}^{+})^{c} & -\frac{1}{\sqrt{2}}(\Psi _{L})^{c}+\frac{1}{\sqrt{6}}\ (%
\tilde{N}_{L})^{c} & (\tilde{E}_{L}^{+})^{c} \\ 
&  &  \\ 
({\Delta }_{L})^{c} & (\tilde{E}_{L}^{-})^{c} & -\frac{2}{\sqrt{6}}\ (\tilde{%
N}_{L})^{c}%
\end{pmatrix}_b\begin{pmatrix}
\frac{s'_1+i a'_1}{\sqrt{2}} \\
 \chi_2^{-} \\
 \frac{w+s_3+i a_3}{\sqrt{2}} \\
\end{pmatrix}.
\end{split}\label{LyOm}
\end{equation}
This product contains the following bilinear terms 
\begin{equation}
\label{eq:8-Yukawa}
\begin{split}
y_{ab}^{\Omega }
\overline{l}_{aL}(\Omega _{bL})^{c}\chi\supset\frac{y_{ab}^{\Omega }w}{\sqrt{2}}\,\left( \overline{%
\nu _{aL}}(\widetilde{\Delta }_{bL})^{c}+\overline{e_{aL}}(\widetilde{E}%
_{bL}^{+})^{c}-\frac{2}{\sqrt{6}}\overline{N_{aL}}(\widetilde{N}%
_{bL})^{c}\right),
\end{split}
\end{equation}
with $a,b=1,2,3$. In turn, the lepton octet mass term in Eq.~(\ref{Ly}) includes the following bilinears
\begin{equation}
\label{eq:8-Mass}   
\frac{M_{8\,ab}}{2} \mathrm{Tr}\left[ \overline{\Omega }_{aL}\left( \Omega _{bL}\right) ^{c}\right]
\supset \frac{M_{8\,ab}}{2} \left[ \left( \overline{\Psi _{aL}}\right) (\Psi
_{bL})^{c}+\overline{\tilde{N}_{aL}}(\tilde{N}_{bL})^{c}+\overline{\tilde{\Delta}
_{aL}}({\Delta }_{bL})^{c}+\overline{\Delta}_{aL}(\tilde{\Delta}%
_{bL})^{c}\right],
\end{equation}
From these expressions we can identify the mass and mixing of the neutral fermions in the model, namely six states
$\Psi_{aL}$, $N_{aL}$, $\tilde{N}_{aL}$, $\nu_{aL}$, $\Delta_{aL}$, and $\widetilde{\Delta }_{aL}$
in each family.

First of all note that, from Eqs.~(\ref{eq:8-Mass}), (\ref{eq:8-Yukawa}), it follows that the three neutral singlet states $\Psi _{a}$ remain unmixed
with other fermions after spontaneous symmetry breaking, their mass matrix being just the bare octet mass $M_{8\,ab}$ in Eq.~(\ref{eq:8-Mass}). 
The two ``dark'' or $M_P$-odd fermions $N_{aL}$ and $\tilde{N}_{aL}$ mix through the mass matrix 
\begin{equation}
M_{N}=\left( 
\begin{array}{cc}
0& M_{N\tilde{N}} \\
M_{N\tilde{N}}^T&M_8
\end{array}%
\right),  \label{MN}
\end{equation}
written in the basis $(N_{L}, \tilde{N}_{L}^{c})$ and with $(M_{N\tilde{N}})_{ab}=-y_{ab}^{\Omega }w/\sqrt{3}$.
Note that this matrix structure resembles the seesaw mechanism though it involves different fields, i.e. heavy electrically neutral Majorana fermions.
Using the general expansion method in \cite{Schechter:1981cv} this matrix can be diagonalized perturbatively by a unitary transformation,
defining six physical heavy Majorana states denoted by $S_{\alpha L}$ through
\begin{equation}
\left( 
\begin{array}{c}
N^c_L \\
\tilde{N}^c_L
\end{array}%
\right)=U S^c_L,  \label{SSt}
\end{equation}
such that $M'_N=U^TM_NU=\mathrm{diag}(M'_\alpha)$, $\alpha=1,\dots,6$. In the following analysis, only the lower blocks of the unitary matrix $U$ will be relevant.
We adopt the following notation for the relation between $N^c_L$ and $S^c_L$:
\begin{equation}
\tilde{N}^c_{aL}=U_{a\alpha} S^c_{\alpha L}. 
\end{equation}
For simplicity, we will assume that the entries of the mass matrix $M_{N\tilde{N}}$ are all real, and consequently that the matrix $U$ is orthogonal.

Finally, Eqs. (\ref{eq:8-Mass})  and   (\ref{eq:8-Yukawa}) provide a tree-level mass matrix for the remaining neutral fermions, including 
the active neutrino fields, which, in the basis $(\nu_{L},\widetilde{\Delta }_{L},\Delta_{L})$, has the form
\begin{equation}
M^{\text{tree}}_{\nu }=\left( 
\begin{array}{ccc}
0 & y^{\Omega }w/\sqrt{2} & 0\\ 
\left( y^{\Omega }\right) ^{T}w/\sqrt{2} & 0 & M_8\\
0 &M_8 &0
\end{array}%
\right).
\end{equation}
It can be shown that this matrix has rank 6, implying that the three light active neutrinos remain massless at tree-level.
This is a very important consistency check of our construction, i.e. neutrinos remain massless at tree-level.
Their non-zero masses arise radiatively at one-loop via the scotogenic mechanism, which we explain in the next subsection.

\subsection{Scotogenic Neutrino masses}

\begin{figure}[tb]
\includegraphics{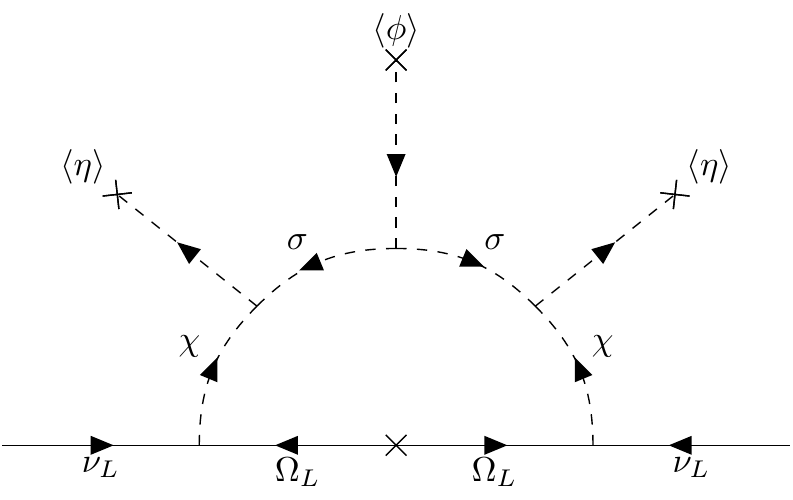} 
\caption{Scotogenic loop for Majorana neutrino mass. }
\label{Fig:Scoto-Nu}
\end{figure}

We first notice that the key ingredient for a scotogenic mechanism for neutrino masses is already present in the basic \3311 framework of Ref.~\cite{CarcamoHernandez:2020ehn}.
Indeed, non-zero neutrino masses arise in our model at one-loop level, from the diagram in Fig.~\ref{Fig:Scoto-Nu}.  
Moreover, two \emph{ad hoc} features of the original scotogenic model in Ref.~\cite{Ma:2006km} become automatic by the 3-3-1-1 embedding:
\begin{enumerate}
\item the discrete symmetry responsible for WIMP dark matter stability is simply matter-parity $M_{P}$ in Eq.(\ref{eq:mp}),
  which survives as a residual discrete symmetry after the complete spontaneous symmetry breaking chain of Eq.(\ref{eq:SSB-chain}). 
\item the dark $\mathrm{SU(2)}$ scalar doublet needed in the scotogenic loop in Fig.~\ref{Fig:Scoto-Nu} is readily identified with the first two components of the $\chi$ triplet,
  whose third component is responsible for $\mathrm{SU(3)_L}$ breakdown.  
\end{enumerate}

Finally, the scotogenic loop is closed in our model by the inclusion of the fermion octet,
which also plays a double role, as mediator of neutrino masses, as well as the dominant field responsible for our proposed gauge coupling unification mechanism, see Sec.~\ref{sec:gauge-coupl-unif}.
\\[-.4cm]

Following from the analysis of the scalar potential in Eq.(\ref{eq:scalar-pot}), the Yukawa interaction terms contained in Eq. (\ref{LyOm}) relevant for the neutrino masses are
\begin{equation}
\begin{split}
y_{ab}^{\Omega }%
\overline{l}_{aL}(\Omega _{bL})^{c}\chi\supset &y_{ab}^{\Omega }
\overline{\nu}_{aL}
\left(
\frac{1}{2}\Psi _{bL}^{c}+\frac{1}{2\sqrt{3}} \tilde{N}_{bL}^{c}\right)
\left(s'_1+i a'_1\right) \\
=&
y_{ab}^{\Omega }
\overline{\nu}_{aL}
\left(
\frac{1}{2}\Psi _{bL}^{c}+\frac{1}{2\sqrt{3}}\ U_{b\alpha}S_{\alpha L}^{c}\right)
\left(U^s_{1i}\varphi_i+i U^a_{1i}\widetilde{\varphi}_i\right).
\end{split}\label{LyOm2}
\end{equation}
Then, according to Fig. \ref{Fig:Scoto-Nu} the one loop level light active neutrino mass matrix is given by 
\begin{equation}
\begin{split}
\left( M_{\nu }\right) _{ab}=&\sum_{c =1}^{3}\sum_{i =1}^{2}\frac{y^{\Omega}_{ac}y^{\Omega}_{bc}M_{8\,c}}{%
32\pi ^{2}}\left[ (U^{s}_{i1})^2\frac{m_{\varphi_{i}}^2}{m_{\varphi_{i}}^{2}-M_{8\,c}^{2}}\ln \left( \frac{m_{\varphi_{i}}^{2}}{%
M_{8\,c}^{2}}\right) -(U^{a}_{i1})^2\frac{m_{\widetilde{\varphi}_i}^2}{m_{\widetilde{\varphi}_i}^{2}-M_{8\,c}^{2}}\ln \left( \frac{m_{\widetilde{\varphi}_i}^{2}}{%
M_{8\,c}^{2}}\right) \right] \\&+\sum_{\alpha =1}^{6}\sum_{i =1}^{2}\frac{(y^{\Omega}U)_{a\alpha}(y^{\Omega}U)_{b\alpha}M'_\alpha}{%
96\pi ^{2}}\left[ (U^{s}_{i1})^2\frac{m_{\varphi_{i}}^2}{m_{\varphi_{i}}^{2}-M'_{\alpha}{}^{2}}\ln \left( \frac{m_{\varphi_{i}}^{2}}{%
M'_{\alpha}{}^{2}}\right) -(U^{a}_{i1})^2\frac{m_{\widetilde{\varphi}_i}^2}{m_{\widetilde{\varphi}_i}^{2}-M'_{\alpha}{}^{2}}\ln \left( \frac{m_{\widetilde{\varphi}_i}^{2}}{%
M'_{\alpha}{}^{2}}\right) \right].
\end{split}
\end{equation}
In the above expression, the mass splitting between the CP even and CP odd scalars running in the internal lines of the loop is induced by the scalar singlet $\sigma$
through the trilinear interactions $\frac{\mu _{s}}{\sqrt{2}} \phi^{\dagger} \sigma \sigma $ and $\frac{\mu _{u}}{\sqrt{2}}\left( \eta ^{\dagger}\chi \right) \sigma $. 
As pointed out before, a degenerate physical scalar spectrum $m_{\varphi _{1,2}}^{2}=m_{\widetilde{\varphi }_{1,2}}^{2}$ is obtained in the limit $\mu _{s},~\mu_{u} \to 0$.
In such limit one therefore obtains vanishing neutrino masses. 
Natural tiny values of the light active neutrino masses are achieved by invoking non-zero couplings $\mu_{s}$ and $\mu_{u}$,
which produce a small mass splitting between the virtual CP-even and CP-odd scalars that take part as mediators in the scotogenic loop. 

Finally, to close this section it is worth mentioning that in the limit $\mu _{s},~\mu_{u} \to 0$, the Lagrangian of the model acquires an accidental $\mathrm{U(1)}$ symmetry,
 under which the scalar fields $\sigma$ and $\phi$ transform with the same charge, whereas the rest of the fields remain invariant.
 This accidental $\mathrm{U(1)}$ symmetry is explicitly broken by the trilinear scalar couplings $\mu_{s}$ and $\mu_{u}$,
 thus implying that in our model the light neutrino masses are symmetry-protected.

\section{WIMP scotogenic dark matter}
\label{sec:wimp-dark-matter}

Due to the presence of a conserved discrete matter-parity surviving spontaneous symmetry breaking, our model contains a potentially stable dark matter candidate,
namely the lightest electrically neutral $M_P$-odd particle. \\[-.4cm] 

The phenomenology of a scalar scotogenic dark matter in our model was previously studied in \cite{CarcamoHernandez:2020ehn}.
There, the viability of the real scalar field $\varphi_2$ as a WIMP dark matter candidate was analyzed within a simplified scenario 
  where all the non-SM fields were assumed to be heavy, with $|\theta_s|\ll 1$, and therefore decoupled.
  In such limit, $\varphi_2$ is mostly composed of the electroweak singlet scalar $\sigma$, and thus it mainly
  annihilates into a pair of Higgs fields, through the simplified Higgs portal quartic scalar interaction $\lambda_{\mathrm{eff}}h^2(\varphi_2)^2$.
  In this regime, the dark matter candidate $\varphi_2$ has a small coupling with the $Z$-boson, evading direct detection Higgs-portal experimental bounds~\cite{XENON:2018voc}.
  
\begin{figure}[btp!]
\centering
\includegraphics[width=0.45\textwidth]{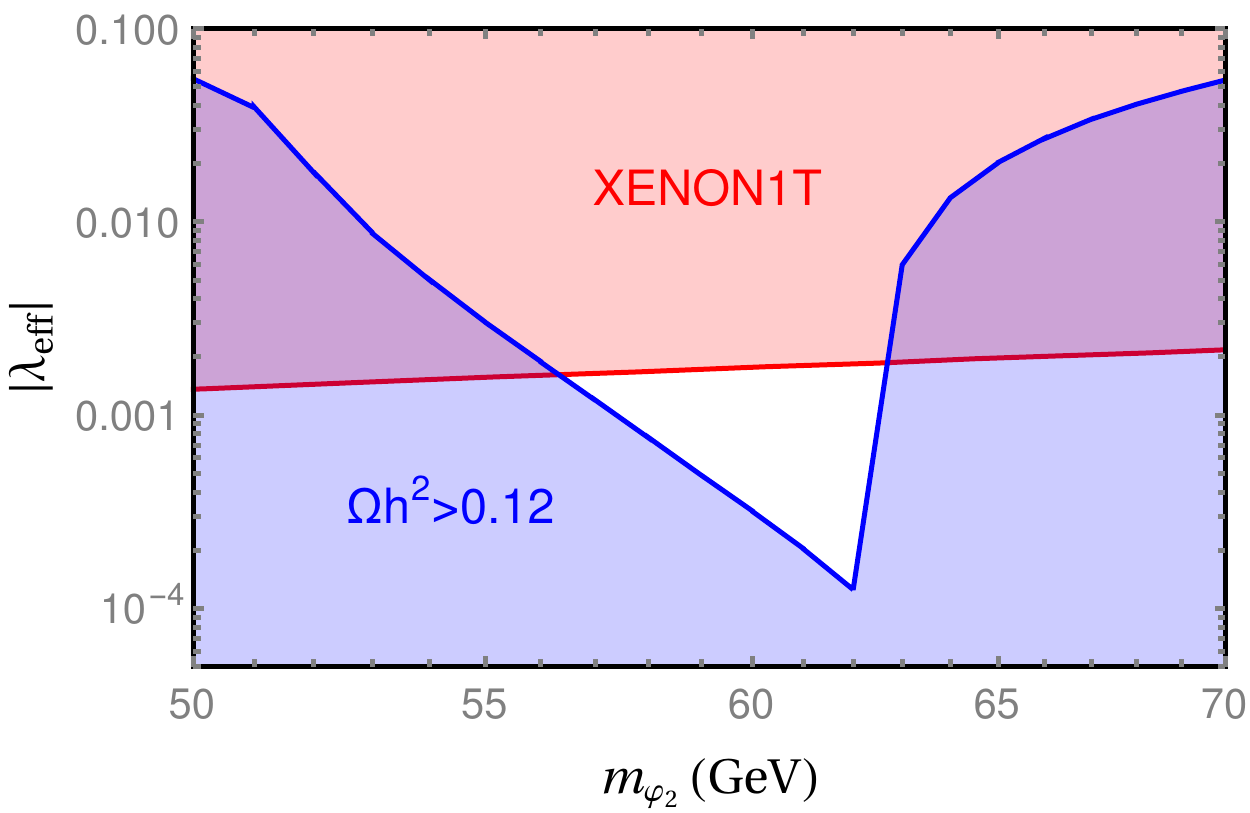} 
\includegraphics[width=0.45\textwidth]{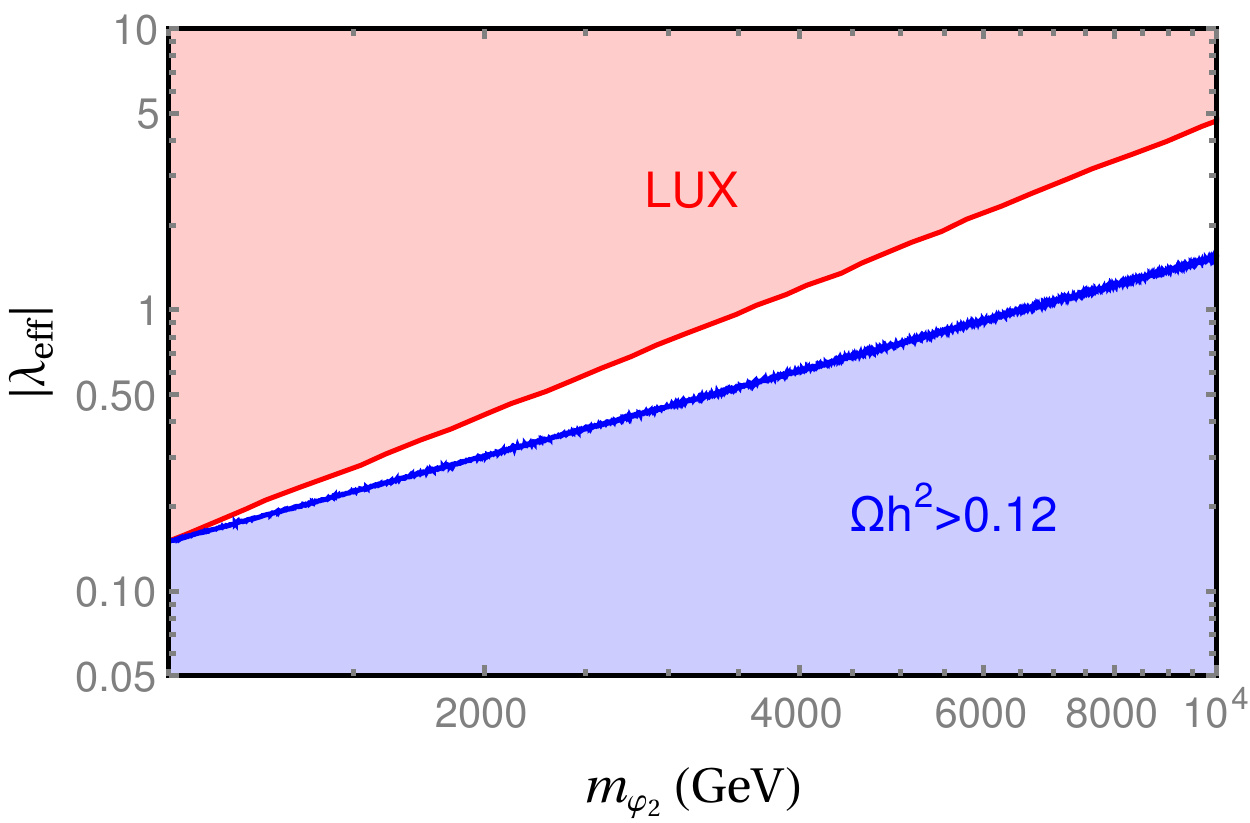}
\caption{Viable mass regions where the field $\varphi_2$ of the simplified model described in the text behaves as a dark matter candidate.
  The red regions correspond to the current direct detection limits~\cite{XENON:2018voc,LUX:2016ggv}.
  The blue regions represent values of the effective coupling $\lambda_{\mathrm{eff}}$ where the corresponding relic density is incompatible with the Planck measurement~\cite{Planck:2018vyg}.
}
\label{fig:low-high}
\end{figure}

Even in such constrained scenario, $\varphi_2$ yields viable relic densities~\cite{CarcamoHernandez:2020ehn}, while evading direct detection bounds.
  This happens in two viable mass regions: near half of the Higgs mass, where resonant annihilation of dark matter into the Higgs boson takes place,
  and also for masses $m_{\varphi_2}$ above $1\,\mathrm{TeV}$, where the direct detection constraints on the effective Higgs portal coupling are weak~\cite{LUX:2016ggv},
 These are shown in the left and right panels in Fig(\ref{fig:low-high}). 
Notice that the allowed regions for scalar dark matter can be considerably larger than those of the simplified scenario described above~\cite{CarcamoHernandez:2020ehn},
due to the presence of re-scattering effects \cite{Kakizaki:2016dza} coming from the other scalar dark fields present in the model.\\[-.4cm]

Concerning fermionic scotogenic dark matter, there are several $M_P$-odd Majorana fermions that may play this role in the present model, 
including the previously discussed fields $\Psi_{aL}$, $N_{aL}$, $\tilde{N}_{aL}$, $\Delta_{aL}$, $\tilde{\Delta}_{aL}$.
As shown above, most of these fieds have masses dictated by the octet mass $M_8$, which plays an important role in gauge coupling unification, as discused in the next section. 
Since we will assume that the scale $M_8$ is higher than the 3-3-1-1 symmetry breaking scale, most of the Majorana fermions will be heavy.
In fact, their masses are restricted by the gauge coupling unification hypothesis, leading to a limited region accomodating viable relic densities. \\[-.4cm]

There are, however, two scenarios for lighter WIMP dark matter fermions. 
The first one emerges from the observation that the fields $N_{aL}$ and $\tilde{N}_{aL}$ undergo a seesaw mechanism when $M_8>M_{N\tilde{N}}$, see Eq.(\ref{MN}).
This can naturally drive the mass of the lightest physical fermion $S_{1L}$ below the few TeV scale.
In this case, the fermionic dark matter candidates can annihilate into a SM fermion pair and neutral lepton pairs via the $s$-channel
  exchange of $Z^{\prime}$ and $Z^{\prime\prime}$ gauge bosons and $t$-channel exchange of neutral scalars, respectively.
  Moreover, a pair of heavy neutral gauge bosons can be produced by the $t$-exchange of Majorana fermions,
  and a scalar pair can be produced via the $t$-exchange of neutral leptons.
 Finally, a neutral scalar and a heavy neutral gauge boson can in turn be created from the $s$ channel exchange of $Z^{\prime}$ and $Z^{\prime\prime}$ of the same neutral fermions.
  Detailed studies of the dark matter relic abundance in the scenario of a Majorana dark matter candidate in radiative neutrino mass models have been carried out in~\cite{Abada:2021yot}.
  It was shown that a fermionic dark matter candidate should be heavier than about $3$ TeV in order to successfully comply with the available experimental constraints.
  Other studies of fermionic dark matter constraints in models with extended gauge symetry were performed in \cite{Huong:2019vej,Alvarado:2021fbw}.
  This $\sim 3$ TeV limit can be qualitatively understood by noticing that the dominant contribution to the annihilation cross section into SM states arises
  from the $s$ channel exchange of heavy neutral gauge bosons $Z^{\prime}$ and $Z^{\prime\prime}$, as pointed out in \cite{Abada:2021yot,Huong:2019vej}. 
  Within the simplified scenario where the $Z^{\prime}$ exchange dominates over the exchange of $Z^{\prime\prime}$ (which corresponds to taking $\Lambda\gg w$),
  the thermally-averaged annihilation cross section of fermionic dark matter candidate into SM states can be estimated as \cite{Huong:2019vej}:
\begin{equation}
\vev{\sigma v}\approx\left(\frac{\alpha}{150\,\text{GeV}}\right)^2\left(\frac{M_{\Omega}}{3\,\text{TeV}}\right)^2\approx\left(\frac{M_{\Omega}}{3\,\text{TeV}}\right)^2\,\text{pb},   
\end{equation}
which shows that the measured value of the dark matter relic abundance
\begin{equation}
\Omega_{DM}h^2=\frac{0.1\,\text{pb}}{\vev{\sigma v}},
\end{equation}
can be naturally explained by having a fermionic dark matter candidate with a mass of the order of $3$ TeV.  \\[-.4cm]

Finally, one can achieve light WIMP dark matter fermions in the present model by giving masses to the $M_P$-odd right-handed neutrinos $\nu_{iR}$
  through spontaneous symmetry breaking with the inclusion of adequate scalar fields.
In such scenario, one can easily accomodate a viable relic density compatible with direct detection bounds.
A dedicated analysis of the WIMP dark matter phenomenology is beyond the scope of this paper. \\[-.4cm] 

\section{Gauge coupling unification}
\label{sec:gauge-coupl-unif}

One of the primary goals of the grand unification programme consists in embedding the SM gauge group ${\cal{G}}_{\rm{SM}}\equiv \mathrm{SU(3)_{c}\otimes SU(2)_{L}\otimes U(1)_{Y}}$
into a bigger gauge group with only one coupling constant.  
At some very high energy scale $M_{U}$, the unified group ${\cal{G}}_{U}$ breaks down to the SM gauge group (or some gauge extension thereof) followed by different evolution of the gauge couplings
leading to their SM values at the electroweak symmetry breaking scale. 
Interestingly, one can have several intermediate scales between the unification scale $M_U$ and the $Z$-pole, $m_Z$, corresponding to multi-stage breaking of ${\cal{G}}_{U}$ to ${\cal{G}}_{\rm{SM}}$. 
The most straightforward route to unification was proposed by Georgi and Glashow \cite{Georgi:1974sy}, who pointed out that the SM can be embedded into the rank-4 simple Lie group $\mathrm{SU(5)}$,
implying the unification of all the SM coupling constants. 

In what follows of this section, instead of specifying a unified gauge group upfront, we will take a phenomenological approach to find the possibility for a dynamical gauge coupling unification.
This exploratory approach not only allows us to unravel the phenomenological possibilities for the unification of the electroweak interaction with the new interactions of our scotogenic framework,
but also provides important hints on the possible interesting UV-completions on which we comment towards the end of this section.  

Having already discussed how the SM gauge group and particle content can be embedded into our 3-3-1-1 theory, in this section we proceed to perform a comprehensive study
of the Renormalisation Group Equations (RGE) in order to determine the conditions under which the SM gauge couplings, embedded into the extended gauge group
$\mathrm{SU(3)_{C}^{}\otimes SU(3)_{L}^{}\otimes U(1)_{X}^{}\otimes U(1)_{N}^{}}$, lead to a successful gauge coupling unification at some high scale $M_{U}$. 
\emph{A priori} we do not make any assumptions about the ultimate unified gauge group. Instead we work with the 3-3-1-1 model, in order to explore various possibilities. 
At the one-loop level the evolution of the gauge couplings $g_{i}$ is governed by the renormalisation group equations (RGEs)
\begin{equation}{\label{gut:1.1}} 
\mu\,\frac{\partial g_{i}}{\partial \mu}=\beta_i(g_i)\equiv \frac{b_i}{16 \pi^2} g^{3}_{i},
\end{equation}
where $\mu$ is the renormalization scale. The behaviour of the gauge couplings with the energy can also be expressed in the form
\begin{equation}{\label{gut:1.2}} 
\frac{1}{\alpha_{i}(\mu_{2})}=\frac{1}{\alpha_{i}(\mu_{1})}-\frac{b_{i}}{2\pi} \ln \left( \frac{\mu_2}{\mu_1}\right),
\end{equation}
where $\alpha_{i}=g_{i}^{2}/4\pi$ and the one-loop beta-coefficients $b_i$ are given by 
\begin{eqnarray}{\label{gut:1.3}} 
	&&b_i= - \frac{11}{3} \mathcal{C}_{2}(G_{i}) 
				 + \frac{2}{3} \,\sum_{R_f} T(R_f) \prod_{j \neq i} d_j(R_f) 
  + \frac{1}{3} \sum_{R_s} T(R_s) \prod_{j \neq i} d_j(R_s).
\label{oneloop_bi}
\end{eqnarray}
Here, $\mathcal{C}_2(G_{i})$ is the quadratic Casimir invariant corresponding to the adjoint representations,
\begin{equation}{\label{gut:1.4}} 
	\mathcal{C}_2(G) \equiv \left\{
	\begin{matrix}
		N & \text{if } \mathrm{SU(N)}, \\
    0 & \hspace{-0.4 cm}\text{if } \mathrm{U(1)}.\end{matrix}
	\right.
\end{equation}
whereas $T(R_f)$ and $T(R_s)$ correspond to the Dynkin indices of the irreducible representation $R_{f,s}$ for a given fermion and scalar, respectively. For the case of $\rm SU(N)$ they are
\begin{equation}{\label{gut:1.5}}  
	T(R_{f,s}) \equiv \left\{
	\begin{matrix}
		1/2 & \text{if } R_{f,s} \text{ is fundamental}, \\
    N   & \hspace{-0.9 cm}\text{if } R_{f,s} \text{ is adjoint}, \\
		0   & \hspace{-0.9 cm}\text{if } R_{f,s} \text{ is singlet}.\end{matrix}
	\right.
\end{equation}
The quantity $d(R_{f,s})$ in (\ref{oneloop_bi}) is the dimension of a given representation $R_{f,s}$ under all gauge groups except for the \mbox{$i$-th}~gauge group under consideration.

As shown above, in our 3-3-1-1 model the electric charge operator is defined as
\begin{equation}{\label{gut:1.6}} 
Q=T_{3}-\frac{1}{\sqrt{3}}T_{8}+X,
\end{equation}
where the $\rm SU(3)_{L}$ generators are normalised as $\rm{Tr}\left(T_{i}T_{j}\right) =\frac{1}{2}\delta _{ij}$. 
Note that the $\mathrm{U(1)_X}$ charge, $X$, enters in the definition of electric charge $Q$ and hence can be related to the hypercharge after the breaking of
$\mathrm{SU(3)_{C}^{}\otimes SU(3)_{L}^{}\otimes U(1)_{X}^{}\otimes U(1)_{N}^{}}$ to the SM gauge group $\mathrm{SU(3)_{C}\otimes SU(2)_{L}\otimes U(1)_{Y}}$ as 
\begin{equation}{\label{gut:1.7}} 
Y=-\frac{1}{\sqrt{3}}T_{8}+X.
\end{equation}
Therefore the initial value for $\alpha_X$ at the 3-3-1-1 symmetry breaking scale $M_X$ can be obtained using the hypercharge $\alpha_Y$ evolution from the $Z$-pole to $M_X$.
In addition, we can define the normalized charge operators $X_{N}$ and $Y_{N}$, which satisfy the relations 
\begin{equation}{\label{gut:1.8}} 
X=n_{X}X_{N},\hspace{1.5cm}Y=n_{Y}Y_{N} \, ,
\end{equation}
with the normalizations of $X$ and the hypercharge $Y$ being related by 
\begin{equation}{\label{gut:1.9}} 
n_{Y}^{2}=\frac{1}{3}+n_{X}^{2}.
\end{equation}
We recall that in an embedding of the SM gauge group into some unified simple group the hypercharge normalization is usually chosen
so that it matches with the normalization for the $\mathrm{SU(N)}$ counterparts in the SM gauge group,
\begin{equation}{\label{gut:1.10}} 
	{\rm{Tr}}[T_i T_j]=\frac{1}{2} \delta_{ij} \, .
\end{equation}
For instance, in a $\mathrm{SU(5)}$ theory by fixing the normalisation of the fundamental representation one can fix the $\mathrm{U(1)_Y}$ normalisation to $n_{Y}^2=5/3$.  
Here, in the absence of any specific unification group we will treat $n_{Y}$, \emph{a priori}, as a free parameter. 

We also notice that the $\mathrm{U(1)_{N}}$ charge does not contribute to the electric charge, and therefore can be considered as an ``electrically neutral new charge'' (ENNC). 
As a result, the initial value for $\alpha_N$ at the 3-3-1-1 symmetry breaking scale $M_X$ and the normalisation of $U(1)_{N}$,
\begin{equation}{\label{gut:1.11}} 
N=n_{N} N_{N}\, ,
\end{equation}
remain free parameters which cannot be fixed by electroweak gauge coupling input values or hypercharge normalization. 
\begin{table}[ht!]
\centering
\begin{tabular}{|c|c|c|c|c|}
\hline
\hspace{0.2cm} Relevant Gauge group \hspace{0.2cm} & Scale of running  \hspace{0.2cm} & \hspace{0.2cm} Gauge group $G_i$ \hspace{0.2cm}
& \hspace{0.2cm}  Notation for $b_i$ \hspace{0.2cm} & \hspace{0.2cm} Value of $b_i$ \hspace{0.2cm}  \\ 
\hline\hline
& & $SU(3)_C$ & $b_{3C} $ &  -7\\ 
$SU(3)_{C}\otimes SU(2)_{L}\otimes U(1)_{Y}$& $M_Z<\mu< M_X$ & $SU(2)_L$ & $b_{2L} $ & $-\frac{19}{6}$\\ 
 & & $U(1)_Y$ & $b_{Y}^{\text{UN}} $ & $\frac{41}{6}$ \\ 
 \hline\hline
& & $SU(3)_C$ & $b_{3C}^X $ &  $-5$\\ 
$SU(3)_{C}^{}\otimes SU(3)_{L}^{}\otimes U(1)_{X}^{}\otimes U(1)_{N}^{}$ & $M_X<\mu< M_8$ & $SU(3)_L$ & $b_{3L}^X $ & $-\frac{13}{2}$\\ 
& & $U(1)_X$ & $b_{X}^{\text{UN}} $ &  $\frac{26}{3}$\\ 
& & $U(1)_N$ & $b_{N}^{\text{UN}} $ &  $\frac{163}{3}$\\ 
  \hline\hline
& & $SU(3)_C$ & $b_{3C}^\Omega =b_{3C}^X$ & $-5$ \\ 
$SU(3)_{C}^{}\otimes SU(3)_{L}^{}\otimes U(1)_{X}^{}\otimes U(1)_{N}^{}$ & $M_8<\mu< M_U$  & $SU(3)_L$ & $b_{3L}^\Omega $ & $-\frac{1}{2}$\\ 
&  & $U(1)_X$ & $b_{X}^{\Omega\,; \text{UN}}= b_{X}^{\text{UN}} $ & $\frac{26}{3}$ \\
&    & $U(1)_N$ & $b_{N}^{\Omega\,;\text{UN}} = b_{N}^{\text{UN}}$ & $\frac{163}{3}$  \\ 
  \hline\hline
\end{tabular}%
\caption{Values of $b_i$ for different gauge groups ($G_i$) relevant for RG running of gauge couplings at different energy scales.}
\label{tab:beta}
\end{table} 

Furthermore, for the sake of generality, we take the 3-3-1-1 symmetry breaking scale ($M_X$) and the fermionic octet mass scale $M_8>M_X$ as independent scales.  
In Table~\ref{tab:beta}, we summarize the one-loop RGE beta coefficients governing the evolution of the relevant gauge couplings at different scales. 
Before addressing the evolution of the $\mathrm{U(1)_{N}}$ gauge coupling, it is straightforward to find the unification scale $M_U$ for $\mathrm{SU(3)_c}$, $\mathrm{SU(2)_L}$ and $\mathrm{U(1)_X}$
and the hypercharge normalization $n_Y$  as a function of the intermediate symmetry breaking scales. First we note that the normalized couplings are related by 
\begin{equation}{\label{gut:1.12}}
n_{Y}^{2} {\left(\alpha^{N}_{Y}\right)}^{-1}=\frac{1}{3}\alpha_{3L}^{-1}+\left(n_{Y}^{2}-\frac{1}{3}\right){\left(\alpha^{N}_{X}\right)}^{-1}. 
\end{equation}
Taking the 3-3-1-1 symmetry breaking scale ($M_X$) and the mass scale for the fermionic octets $\Omega_{aL}$, $a=1,2,3$, $M_8>M_X$ as independent parameters, and using Eq. \eqref{gut:1.2} we
then obtain 
\begin{eqnarray}
\alpha^{-1}_{U} &=& \frac{1}{n_{Y}^{2}-\frac{1}{3}} 
\left\{
	\alpha_{\text{em}}^{-1}(M_{Z})\cos^{2}\theta_{w} (M_{Z})
	-\frac{1}{3}\alpha^{-1}_{2L}(M_Z)
	-\frac{b^{\text{UN}}_{Y}-\frac{1}{3}b_{2L}}{2\pi}
	 \ln\left(\frac{M_X}{M_Z}\right)
	-\frac{b^{\text{UN}}_{X}}{2\pi}
	 \ln\left(\frac{M_U}{M_X}\right)
\right\},{\label{gut:1.13}} \\
\alpha^{-1}_{U}&=&\alpha^{-1}_{2L}(M_Z)-\frac{b_{2L}}{2\pi}\ln\left(\frac{M_X}{M_Z}\right)-\frac{b_{3L}^X}{2\pi}\ln\left(\frac{M_8}{M_X}\right)-\frac{b_{3L}^\Omega}{2\pi}\ln\left(\frac{M_U}{M_8}\right),{\label{gut:1.14}}\\
\alpha^{-1}_{U}&=&\alpha^{-1}_{3C}(M_Z)-\frac{b_{3C}}{2\pi}\ln\left(\frac{M_X}{M_Z}\right)-\frac{b^{X}_{3C}}{2\pi}\ln\left(\frac{M_U}{M_X}\right)\, , {\label{gut:1.15}}
\end{eqnarray}
where we note that the fermionic octets only affect the evolution of $\alpha_{3L}$ from $M_8$ to $M_U$. 
Hence in the above we denote the beta coefficients for the running of $\mathrm{SU(3)_L}$ from $M_X$ to $M_8$ by $b^{X}_{3L}$ and $M_8$ to $M_U$ by $b^{\Omega}_{3L}$, respectively.
From Eqs. (\ref{gut:1.14}) and (\ref{gut:1.15}) one obtains the unification scale $M_U$ as a function of $M_{X}$ and $M_{8}$ as 
\begin{equation}{\label{gut:1.16}}
M_U(M_X,M_8)=\frac{M_{X}^{\frac{b_{3C}^X}{b_{3C}^X-b_{3L}^\Omega}}}{M_{8}^{\frac{b_{3L}^\Omega}{b_{3C}^X-b_{3L}^\Omega}}}\left(\frac{M_8}{M_X}\right)^{\frac{b_{3L}^X}{b^{X}_{3C}-b_{3L}^\Omega}} \left(\frac{M_X}{M_Z}\right)^{\frac{b_{2L}-b_{3C}}{b^{X}_{3C}-b_{3L}^\Omega}} \exp \left[2\pi \frac{\alpha^{-1}_{3C}(M_Z)-\alpha^{-1}_{2L}(M_Z)}{b^{X}_{3C}-b_{3L}}\right].
\end{equation}
Using Eqs. (\ref{gut:1.12}) and (\ref{gut:1.13}) the normalization $n_Y^2$ can be obtained as
\begin{eqnarray}{\label{gut:1.17}}
n_{Y}^{2}&=&\frac{1}{3}+\left[\alpha_{\text{em}}^{-1}(M_{Z})\cos^{2}\theta_{w} (M_{Z})-\frac{1}{3}\alpha^{-1}_{2L}(M_Z)-\frac{b^{\text{UN}}_{Y}-\frac{1}{3}b_{2L}}{2\pi}\ln\left(\frac{M_X}{M_Z}\right)+  \frac{b^{\text{UN}}_{X}}{2\pi}\ln\left( \frac{M_U(M_X,M_8)}{M_X}\right)\right]\nonumber\\
&\times& \left[\alpha^{-1}_{2L}(M_Z)-\frac{b_{2L}}{2\pi}\ln\left(\frac{M_X}{M_Z}\right)-\frac{b_{3L}^X}{2\pi}\ln\left(\frac{M_8}{M_X}\right)-\frac{b_{3L}^\Omega}{2\pi}\ln\left(\frac{M_U(M_X,M_8)}{M_8}\right)\right]^{-1} \, ,\nonumber\\
\end{eqnarray}
where $M_U(M_X,M_8)$ is given by Eq. (\ref{gut:1.16}).  The one-loop beta coefficients relevant for the running between different scales are collected in Table \ref{tab:beta}.

\begin{figure}[btp!]
\centering
\includegraphics[width=0.45\textwidth]{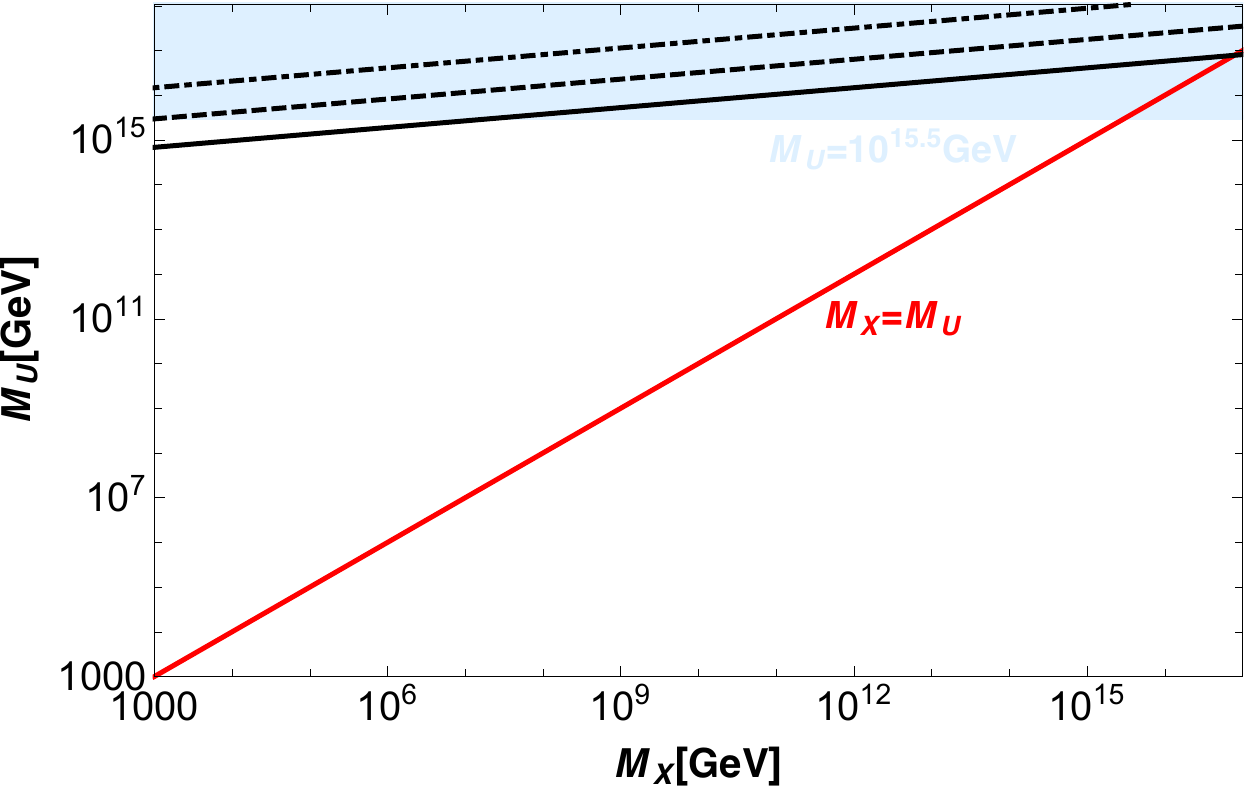}
\includegraphics[width=0.45\textwidth]{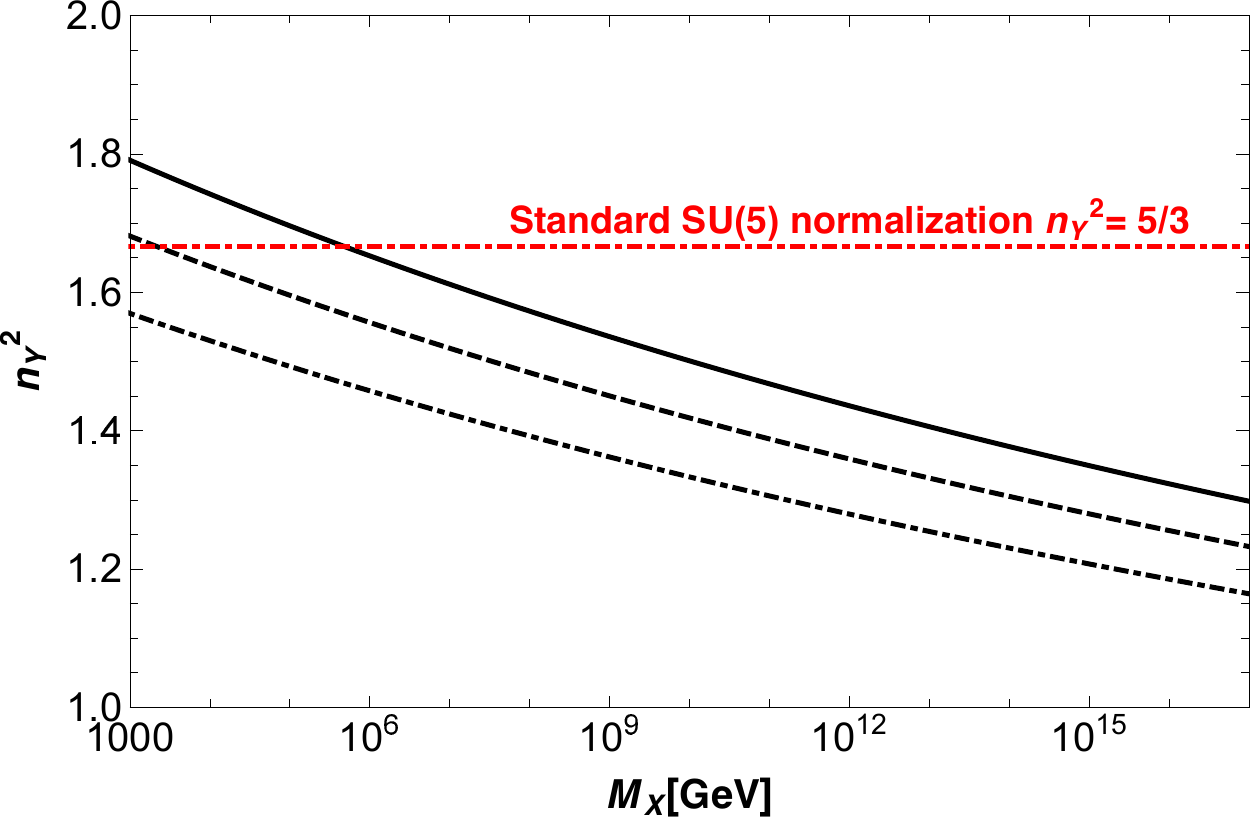}\newline
\caption{(Left) unification scale $M_U$ as a function of the 3-3-1-1 symmetry breaking scale $M_X$, for three benchmark choices $M_8=M_X$ (solid curve),
  $M_8=3 M_X$ (dashed curve) and $M_8=10 M_X$ (dot-dashed curve).   (Right) hypercharge normalization $n_Y^2$ as a function of $M_X$, for the same benchmark choices as the left panel.}
\label{fig:gut1}
\end{figure}

In Fig.~\ref{fig:gut1} left plot, we show the unification scale as a function of the 3-3-3-1 symmetry breaking scale $M_X$ (c.f. Eq. (\ref{gut:1.16})),
for three different benchmark choices $M_8=M_X$ (solid curve), $M_8=3 M_X$ (dashed curve) and $M_8=10 M_X$ (dot-dashed curve).
The blue band indicates the range for the unification scale consistent with the current experimental limit on proton decay lifetime if the 3-3-1-1 gauge group is embedded in a unified gauge group.
However, we note that a dynamical gauge coupling unification achieved with an anomaly-free set (under the 3-3-1-1 gauge group) of
fields \cite{Eichten:1982pn,Fishbane:1983hf,Fishbane:1984zv,Foot:1988qx,Frampton:1993bp,Batra:2005rh,Fonseca:2015aoa}, is not subject to such a constraint.
The tilted red line corresponds to the asymptotic limit $M_X=M_U$.
In Fig.~\ref{fig:gut1} right plot, we show the corresponding hypercharge normalization as a function of the 3-3-1-1 symmetry breaking scale $M_X$, for the benchmark choices described above.
The red line shows the standard hypercharge normalization for a reference $\mathrm{SU(5)}$ unified theory.  
Note that for a given benchmark $M_8$ value, the relevant (solid, dashed or dot-dashed) curves in the left and right panels of Fig.~\ref{fig:gut1} correspond 
to the $\mathrm{SU(3)_c \times SU(3)_L \times U(1)_X}$ unification scale $M_U$ and the relevant hypercharge normalization required for a successful unification of gauge couplings.
Therefore, such a curve represents a family of gauge coupling unification possibilities, with each point corresponding to a particular choice of the 3-3-1-1 symmetry breaking scale $M_X$.  

\begin{figure}[tbp!]
\centering
\includegraphics[width=0.45\textwidth]{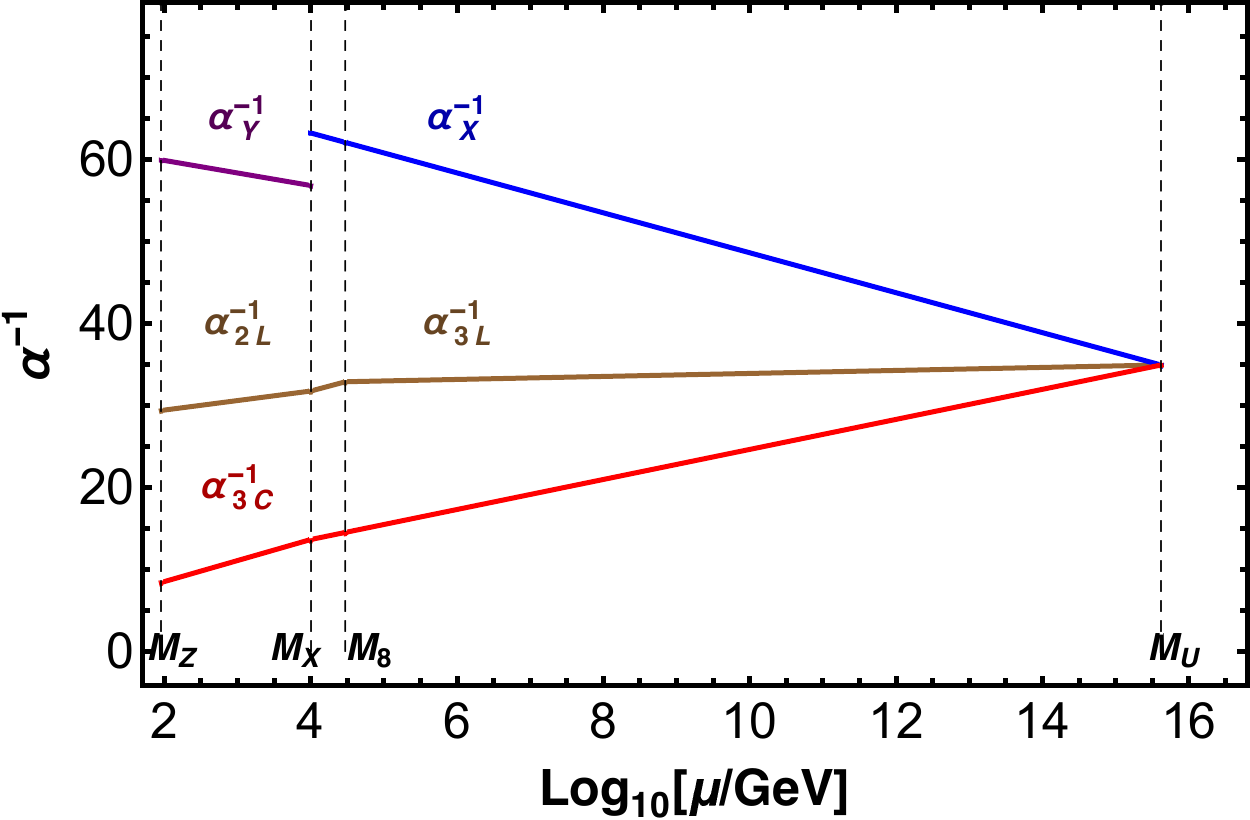}
\caption{An example of $\mathrm{SU(3)_c \times SU(3)_L \times U(1)_X}$ unification for the 3-3-1-1 symmetry breaking scale $M_X=10$ TeV and $M_8=3 M_X=30$ TeV,
  corresponding to the dashed curves in Fig.~\ref{fig:gut1}.}
\label{fig:gut2}
\end{figure}

In Fig.~\ref{fig:gut2} we show such an example point in the dashed curves in Fig.~\ref{fig:gut1} with the 3-3-1-1 symmetry breaking scale $M_X=10$ TeV and $M_8=3 M_X=30$ TeV,
demonstrating a successful $\mathrm{SU(3)_c \times SU(3)_L \times U(1)_X}$ unification. 

Having discussed the $\mathrm{SU(3)_c \times SU(3)_L \times U(1)_X}$ unification, we now explore the possibility of $\mathrm{U(1)_{N}}$ unification with $\mathrm{SU(3)_c \times SU(3)_L \times U(1)_X}$.
First we note that, \emph{a priori}, it is a valid theoretical possibility that $\mathrm{SU(3)_c \times SU(3)_L \times U(1)_X}$ can first unify into a larger gauge group $G_{3-3-1}$,
independent of $\mathrm{U(1)_{N}}$, and at some higher energy scale the unification of $\mathrm{U(1)_{N}}$ and $G_{3-3-1}$ takes place. 
However, for the sake of simplicity, we will only consider the case of $\mathrm{U(1)_{N}}$ unifying at the same scale of $G_{3-3-1}$ unification.

\begin{figure}[btp!]
\centering
\includegraphics[width=0.45\textwidth]{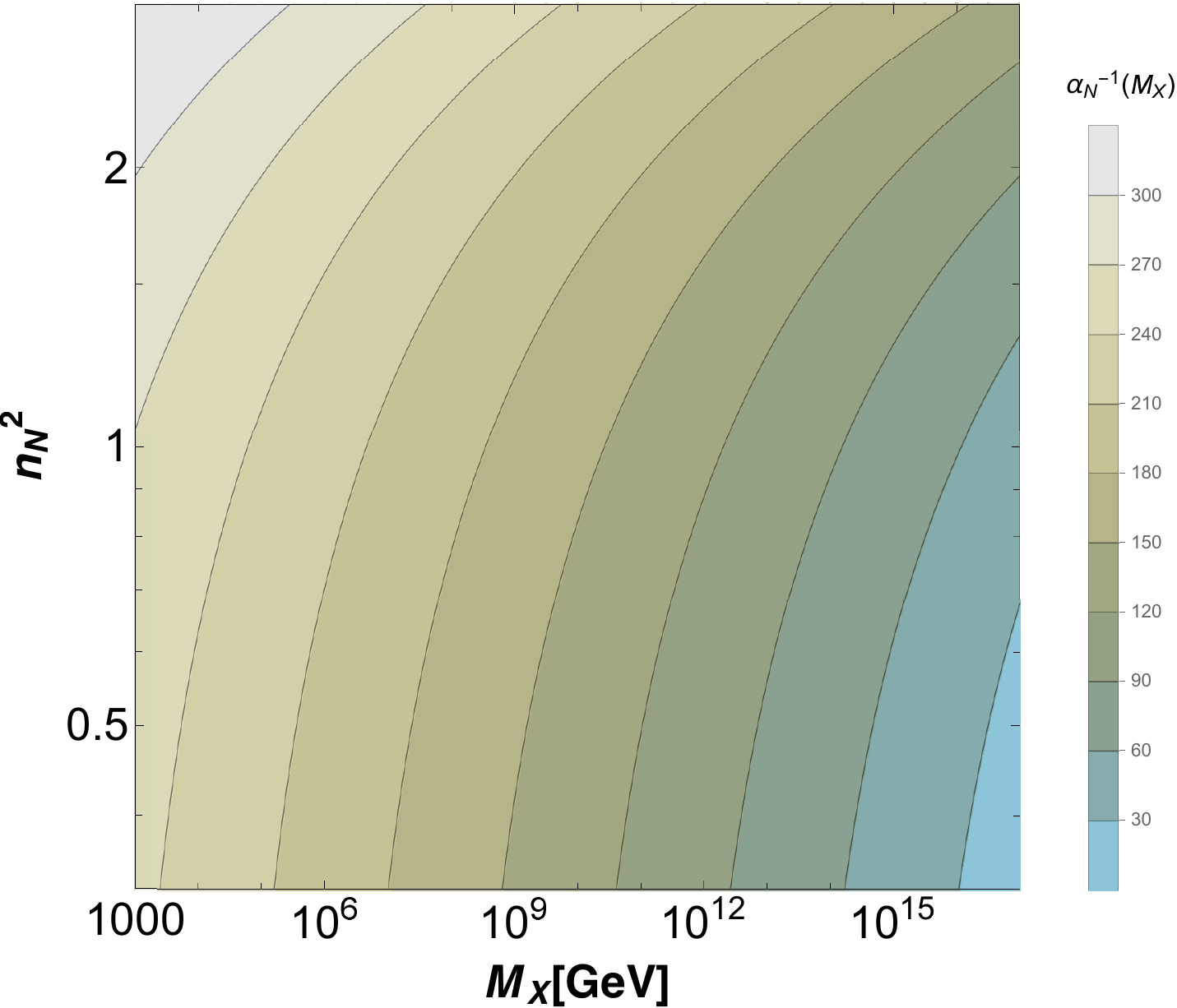}
\includegraphics[width=0.45\textwidth]{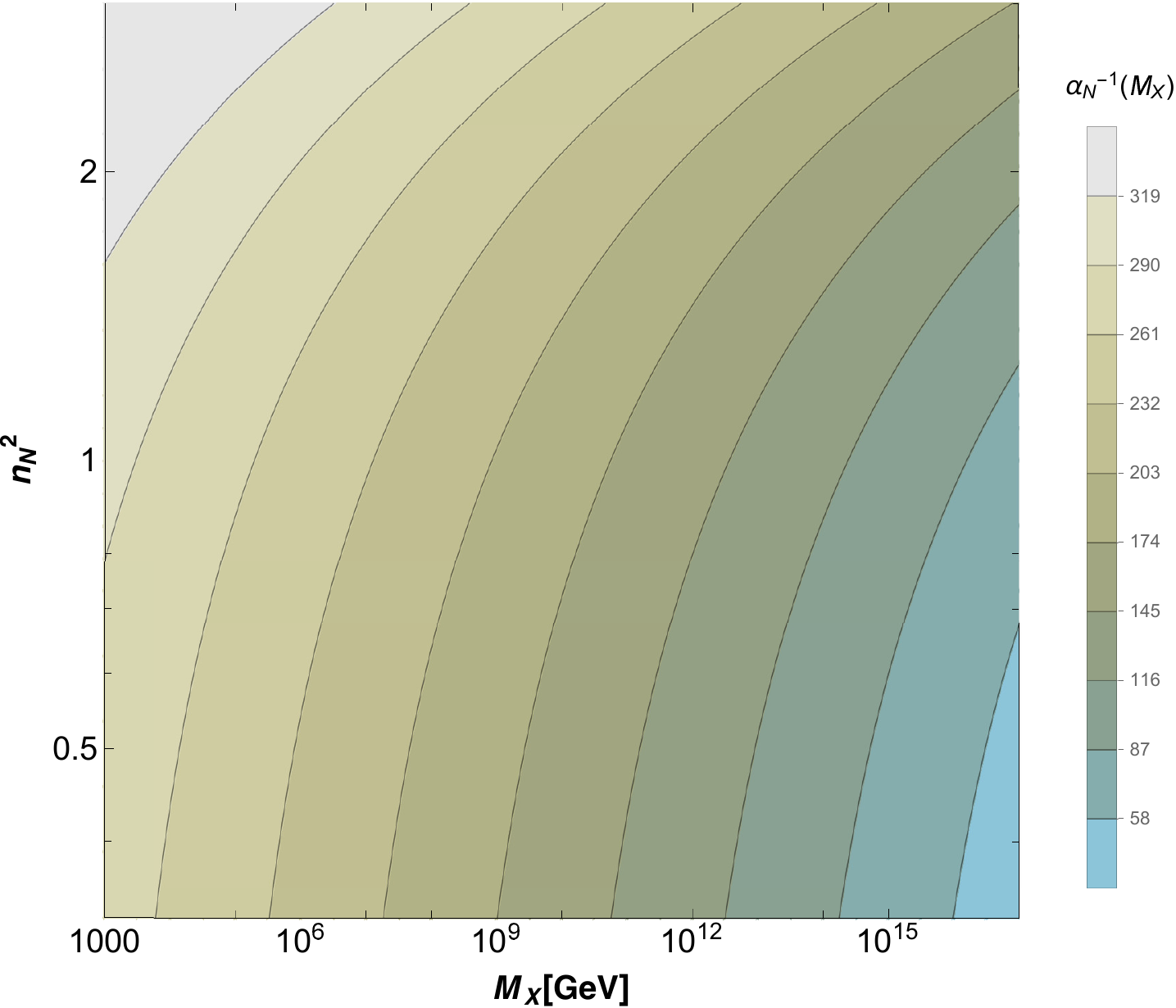}\newline
\caption{\dgreen The left panel shows the $\alpha_N^{-1}$ contours in the $\mathrm{U(1)_N}$ normalisation ($n_N^2$) vs the 3-3-1-1 symmetry breaking scale $M_X$ plane,
  for a benchmark choice $M_8=M_X$, while the right one is the same as the left, but with $M_8= 10 M_X$.  }
\label{fig:gut3}
\end{figure}
We recall that $\mathrm{U(1)_{N}}$ does not contribute to the electric charge and therefore the initial value for $\alpha_N$ at $M_X$ and its normalization $n_{N}$ defined in Eq.~\eqref{gut:1.8}
are not fixed by the hypercharge and electroweak input parameters.  
Therefore, in order to derive the viable initial value for $\alpha_N$ at the 3-3-1-1 symmetry breaking scale $M_X$ and the normalisation $n_{N}$, we require that the $\mathrm{U(1)_{N}}$ coupling  
must unify with the remaining gauge groups associated to $\mathrm{SU(3)_{C}}$, $\mathrm{SU(3)_{L}}$, $\mathrm{U(1)_{X}}$ at the same scale $M_{U}$, leading to the relation 
\begin{equation}{\label{gut:1.18}}
\alpha_U^{-1}=\alpha _{3C}^{-1}(M_{U})=\alpha _{3L}^{-1}(M_{U})=\left( \alpha
_{X}^{N}(M_{U})\right) ^{-1}=\left( \alpha _{N}^{N}(M_{U})\right) ^{-1}, 
\end{equation}
subject to the $b$ coefficients shown in Table \ref{tab:beta}. This in turn yields
\begin{equation}{\label{gut:1.19}}
\alpha_N^{-1}(M_X)=n_N^2 \alpha_U^{-1} + \frac{b_N^{\text{UN}}}{2 \pi}\ln\left( \frac{M_U}{M_X}\right)\,,
\end{equation}
where $\alpha_U^{-1}$ is obtained using any of the Eqs. \eqref{gut:1.10}, \eqref{gut:1.11}, or \eqref{gut:1.12}, and  $M_U(M_X,M_8)$ is given by Eq. (\ref{gut:1.14}). 
In Fig. \ref{fig:gut3} we show the contours for $\alpha_N^{-1}(M_X)$ in the $U(1)_N$ normalisation($n_N^2$) vs 3-3-1-1 symmetry breaking scale $M_X$ plane
(predicted by the requirement that the $\mathrm{U(1)_{N}}$ coupling must unify with the remaining gauge groups associated to $\mathrm{SU(3)_{C}}$, $\mathrm{SU(3)_{L}}$, $\mathrm{U(1)_{X}}$
at the same scale $M_{U}$), for two benchmark choices: in the left plot $M_8=M_X$ and in the right one $M_8= 10 M_X$. 

\begin{figure}[tbp!]
\centering
\includegraphics[width=0.45\textwidth]{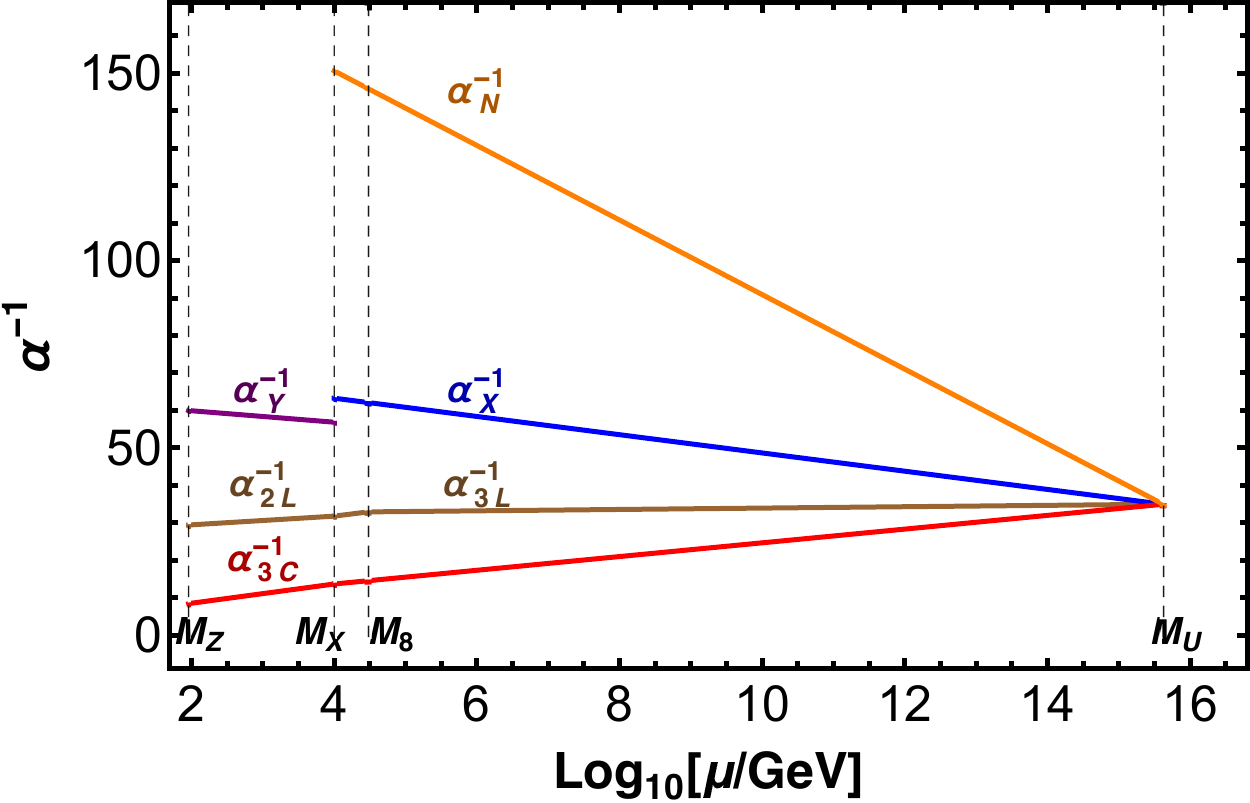}
\caption{An example of $\mathrm{SU(3)_c \times SU(3)_L \times U(1)_X\times U(1)_N}$ unification for a phenomenologically accessible 3-3-1-1 symmetry breaking scale $M_X=10$ TeV
  and $M_8=3 M_X=30$ TeV, corresponding to the dashed curves in Fig.~\ref{fig:gut1}.}
\label{fig:gut4}
\end{figure}

In order to explicitly give an example of a unification scenario, in Fig \ref{fig:gut4} we show $\mathrm{SU(3)_c \otimes SU(3)_L \otimes U(1)_X\otimes U(1)_N}$ unification
for a 3-3-1-1 symmetry breaking scale $M_X=10$ TeV and $M_8=3 M_X=30$ TeV, with the corresponding input for $\alpha_N^{-1}(M_X)$ computed using Eq. \eqref{gut:1.16}.  

\vskip .2cm

Intriguingly, we notice from Figs.~\ref{fig:gut1} and \ref{fig:gut3} that successful gauge coupling unification can occur for a 3-3-1-1 symmetry breaking scale $M_X$
and fermionic octet mass scale $M_8$ around  $\mathcal{O}$(10) TeV, accessible at the current and future collider experiments.
Therefore, our 3-3-1-1 model provides a very exciting phenomenological alternative for having new physics at an energy scale around $\mathcal{O}$(10) TeV
in the form of the gauge bosons associated with 3-3-1-1 symmetry breaking, as well as the fermionic octet.
Such mass scales can not only be explored at collider experiments~\cite{Deppisch:2013cya}, but also lead to interesting charged lepton flavour violation signals~\cite{Boucenna:2015zwa,CarcamoHernandez:2019vih,Hue:2021xap,Hung:2021fzb}.
Moreover they may also be probed in low-energy neutrino experiments, e.g. neutrinoless double beta decay searches~\cite{Santos:2017jbv}.

To conclude this section we comment on the possible embeddings of the 3-3-1-1 gauge group.  
One of the minimal possibilities is to unify the $\mathrm{SU(3)_c \otimes SU(3)_L \otimes U(1)_X\otimes U(1)_N}$ gauge group inside $\mathrm{SU(6) \otimes U(1)_N}$. 
In this case the $\mathrm{SU(3)_c \otimes SU(3)_L \otimes U(1)_X}$ part of the 3-3-1-1 gauge group unifies into $\mathrm{SU(6)}$~\cite{Deppisch:2016jzl,Li:2019qxy}. 
The fermion content of the model (except for the octets) can be embedded into the anomaly-free combination of $\mathrm{SU(6)}$ representations 
$\bar{6}+\bar{6}+15+20$, while the octets can be embedded in the $35$ multiplet of $\mathrm{SU(6)}$ which does not contribute to the anomaly. 
Given that the specific multiplicity of the fermionic triplets in the 3-3-1-1 model is dictated by the number of families,
the required combination of anomaly-free $\mathrm{SU(6)}$ multiplets to accomodate such structure requires additional fields. 
In order to identify the required multiplets in the unified theory it may be useful to make use of flux breaking tools implemented through the Hosotani mechanism \cite{Hosotani:1983xw}.
As seen above, by normalizing the fundamental representation of $\mathrm{SU(6)}$ the hypercharge and $\mathrm{U(1)_X}$ normalizations can be fixed to $n_Y=\sqrt{5/3}$ and $n_X=2/\sqrt{3}$ respectively.
Moreover, the $\mathrm{SU(6)}$ multiplets required for 3-3-1-1 unification can be further embedded in a $\mathrm{E(6)}$ theory with one of its maximal subgroups being $\mathrm{SU(6)\otimes SU(2)}$.
The 27 representation of $\mathrm{E(6)}$ can break into $\bar{6}$ and $15$ representations of $\mathrm{SU(6)}$ and the 78 can break into $35$, $20$ and $1$ representations of $\mathrm{SU(6)}$. 
The $\mathrm{E(6)}$ embedding can be particularly interesting from the perspective of $\mathrm{E(6)}$ $F$-theories~\cite{Gursey:1975ki,Beasley:2008kw,King:2010mq,Callaghan:2011jj,Callaghan:2013kaa}.
Finally, note that a 3-3-1-1  embedding into a unified $\mathrm{SU(6)\otimes U(1)_N}$ group would lead to further constraints on the unification scale,
due to the fact that the $\mathrm{SU(6)}$ gauge bosons can mediate a proton decay mode such as $p\to e^+ \pi^0$.
Experimental searches for the latter lead to a stringent limit $M_U\gtrsim 10^{15.5}$ GeV~\cite{ParticleDataGroup:2020ssz}.


\section{Summary and conclusions}
\label{sec:Conclusions} 

 As a follow-up of our previous work we have now proposed a scotogenic scheme where dark matter stability is ensured by a gauged matter parity symmetry, Eq.~(\ref{eq:mp}).
 The same physics responsible for neutrino mass generation drives the unification of the fundamental gauge couplings.
  A crucial role is played by the leptonic octets in the model: they are responsible for generating the light active neutrino masses through a Scotogenic mechanism (see Fig.~\ref{Fig:Scoto-Nu}),
  while driving gauge coupling unification (see Figs.~\ref{fig:gut1}-\ref{fig:gut4}). Their masses can be accessible to experiments at $\mathcal{O}$(10) TeV scale. 
 Taking such dynamical unification approach as the guiding principle, we have used the unification of the electrically neutral $\mathrm{U(1)_N}$ symmetry
 with $\mathrm{SU(3)_c \otimes SU(3)_L \otimes U(1)_X}$ to predict the initial value and normalization for the coupling strength of the new interaction
 associated with $\mathrm{U(1)_N}$  (see Fig.~\ref{fig:gut3}).
 Indeed, while not exclusive, this approach is very attractive, as it naturally predicts the free parameters associated with the new interaction. 
 The construction is suggestive of a \emph{plethora} of new physics associated to the new gauge bosons and to the exotic states dictated by the 3-3-1-1 gauge symmetry,
 which can be probed at future collider and low-energy experiments. \\[-.4cm] 
 
 Note that our model does not require supersymmetry, though it can me made supersymmetric, should one desire that route to address the gauge hierarchy problem
 instead of, say, the warping of extra dimensions.
 In contrast, our suggestion provides a potentially testable approach where other drawbacks of the SM are addressed in an interconnected manner, such as
 \begin{itemize}
 \item number of the fermion families equals the number of colors,
 \item WIMP dark matter mediates neutrino mass generation,
 \item dark matter stability results from a residual gauge matter-parity,
 \item dynamical unification of gauge couplings.
 \end{itemize}
In short, we have illustrated an idea which seems worth of a dedicated scrutiny of its potential implications.

\acknowledgements 
\noindent

Work supported by the Spanish grants PID2020-113775GB-I00 (AEI / 10.13039/501100011033) and PROMETEO/2018/165 (Generalitat Valenciana).
A.E.C.H and S.K. are supported by ANID-Chile FONDECYT 1210378 and ANID-Chile FONDECYT 1190845 as well as by ANID PIA/APOYO AFB180002 and Milenio-ANID-ICN2019\_044.
C.H. acknowledges support from the DFG Emmy Noether Grant No. HA 8555/1-1. CAV-A is supported by the Mexican Catedras CONACYT project 749 and SNI 58928. The relic abundance and
direct detection constraints were calculated using the MicroOmegas package \cite{Belanger:2018ccd} at GuaCAL (Guanajuato Computational Astroparticle Lab).

\appendix

\bibliographystyle{utphys}
\bibliography{bibliography}

\end{document}